\documentclass[a4paper,fleqn]{cas-sc}

\usepackage[authoryear,longnamesfirst]{natbib}
\usepackage[right]{lineno}
\usepackage{paralist} 
\usepackage{multirow}
\usepackage{threeparttable}
\usepackage{multicol}
\usepackage{stfloats}
\usepackage{float}
\usepackage{enumerate}
\usepackage{makecell}
\usepackage{rotating}
\usepackage{graphicx}
\usepackage{placeins} 
\usepackage{booktabs} 
\usepackage{diagbox}
\usepackage{siunitx} 
\usepackage{subcaption}
\usepackage{arydshln} 
\usepackage{dcolumn} 
\usepackage{longtable} 
\usepackage{setspace}
\newcolumntype{d}[1]{D{.}{.}{#1}} 
\def\tsc#1{\csdef{#1}{\textsc{\lowercase{#1}}\xspace}}
\tsc{WGM}
\tsc{QE}
\tsc{EP}
\tsc{PMS}
\tsc{BEC}
\tsc{DE}

\begin{document}
\doublespacing
\let\WriteBookmarks\relax
\def\floatpagepagefraction{1}
\def\textpagefraction{.001}

\shorttitle{Multidimensional Assessment of Takeover Performance}
\shortauthors{K. Liang et~al.}

\title [mode = title]{Multidimensional Assessment of Takeover Performance in Conditionally Automated Driving}      




%
\author{Kexin Liang}[orcid=0000-0002-8417-3974]
\cormark[1]

\fnmark[a]

\ead{K.Liang-4@tudelft.nl}



\affiliation[a]{organization={Transport and Planning Department, Faculty of Civil Engineering and Geosciences, Delft University of Technology},
    country={The Netherlands}}

\affiliation[b]{organization={Centre for Transport Studies, Department of Civil, Environmental and Geomatic Engineering, University College London},
    country={UK}}

\author{ Jan Luca Kästle}
\fnmark[b]

\author{ Bani Anvari}
\fnmark[b]
    
\author{ Simeon C. Calvert}
\fnmark[a]

\author{ J.W.C. van Lint}
\fnmark[a]






\cortext[cor1]{Corresponding author at: Transport and Planning Department, Faculty of Civil Engineering and Geosciences, Delft University of Technology.}



\begin{abstract}
When automated driving systems encounter complex situations beyond their operational capabilities, they issue takeover requests, prompting drivers to resume vehicle control and return to the driving loop as a critical safety backup. However, this control transition places significant demands on drivers, requiring them to promptly respond to takeover requests while executing high-quality interventions. To ensure safe and comfortable control transitions, it is essential to develop a deep understanding of the key factors influencing various takeover performance aspects. This study evaluates drivers’ takeover performance across three dimensions: response efficiency, user experience, and driving safety— using a driving simulator experiment. EXtreme Gradient Boosting (XGBoost) models are used to investigate the contributions of two critical factors, i.e., Situational Awareness (SA) and Spare Capacity (SC), in predicting various takeover performance metrics by comparing the predictive results to the baseline models that rely solely on basic Driver Characteristics (DC). The results reveal that \begin{inparaenum}[(i)]
   \item  higher SA enables drivers to respond to takeover requests more quickly, particularly for reflexive responses; and
   \item SC shows a greater overall impact on takeover quality than SA, where higher SC generally leads to enhanced subjective rating scores and objective execution trajectories.
\end{inparaenum} These findings highlight the distinct yet complementary roles of SA and SC in shaping performance components, offering valuable insights for optimizing human-vehicle interactions and enhancing automated driving system design.
\end{abstract}










\maketitle

\section{Introduction}

Automated driving systems continue to advance but still face limitations in handling complex and risky driving scenarios. When these systems reach their operational boundaries, in many cases they issue a takeover request, prompting drivers to resume manual control of the vehicle within constrained time budgets. Ensuring smooth vehicle control transitions is crucial for avoiding potential hazards and providing a comfortable driving experience.

Control transitions are challenging because they require drivers to timely shift from a state of passive monitoring or non-driving activities to active vehicle control \citep{hu2023challenges, lu2016human}. This shift is not instantaneous; it demands time and cognitive resources to assess the situation, make decisions, and act. If drivers do not take over vehicle control promptly enough or if their takeover responses are poor in quality, the risks of accidents increase, and the overall user experience deteriorates. Thus, understanding the factors that influence drivers' takeover performance during control transitions and accordingly adopting necessary interventions are essential for enhancing the safety and comfort of conditionally automated driving.

A widely acknowledged factor influencing drivers' takeover performance is Situational Awareness (SA), which reflects how well a driver perceives, understands, and projects the surrounding environment \citep{endsley2020situation, van2018generic}. Studies by \cite{van2013retrieving} and \cite{tan2022effects} demonstrated that drivers with higher SA levels had greater success in executing safe takeover maneuvers. \cite{vlakveld2018situation} found that drivers' SA was closely related to their hazard perception, suggesting that drivers with better SA are more capable of recognizing potential risks, and consequently, reducing crash rates. \cite{mckerral2023supervising} pointed out that improved SA enhanced takeover performance by guiding drivers' responses to takeover requests. Although the positive relationship between SA and takeover performance is well-documented, the concept of ``takeover performance'' is multifaceted and encompasses various aspects of driver behaviours and responses during control transitions \citep{li2023human}. Notably, certain aspects of takeover performance can even conflict: for instance, a shorter takeover time does not always indicate better quality \citep{wu2022does, wang2022shorter}. While a shorter time to resume meaningful human control of the vehicle \citep{calvert2024designing} refers to a more efficient takeover, it may come at the cost of comfort if achieved through abrupt actions, such as hard braking, and can also pose pressure on following vehicles to respond effectively which increases crash risks \citep{wu2022does}. Furthermore, discrepancies between drivers' subjective evaluations of their takeover quality and objective metrics based on driving trajectory have been observed \citep{guo2021effects}, though not well studied. These complexities underscore the need for a more detailed understanding of how SA affects various dimensions of takeover performance.

In addition to SA, this study investigates another critical factor influencing takeover performance: Spare Capacity (SC), derived from Task-Capability Interface (TCI) theory \citep{fuller2005towards}. TCI theory suggests that drivers adjust their behaviours based on the dynamic interactions between their perceived task capability (TC) and task demands (TD) to maintain their perceived safety margin at an acceptable level \citep{fuller2005towards}. On this basis, SC is defined as TC minus TD \citep{fuller2000task}, inversely capturing how difficult a scenario is for a driver to manage. As \cite{fuller2011driver} noted, drivers with lower SC were more vulnerable to performance errors and the challenges of high-demand situations. In the context of automated driving, \cite{muller2021effects} used drivers' secondary task performance as an indicator of SC, finding that drivers engaged in high-workload secondary tasks exhibited reduced SC and required longer reaction times. Similarly, our previous work \citep{liang2024examining} found that drivers with higher SC levels during takeover tasks generally resumed conscious manual control of the vehicle more quickly. We argue that SC has the potential to influence other aspects of takeover performance, especially takeover qualities, but its impacts require further investigation.

This study aims to systematically investigate the effects of SA and SC on various aspects of takeover performance. A driving simulator experiment is conducted, where drivers are instructed to take over vehicle control from automation across nine scenarios (three traffic densities $*$ three non-driving related tasks). For each takeover, drivers’ SA and SC are measured using both \begin{inparaenum}[(i)]
    \item questionnaires that are adopted from well-established instruments for takeover contexts, and
    \item cognition-related eye movement metrics. 
\end{inparaenum} Meanwhile, takeover performance is systematically evaluated along three dimensions: \begin{inparaenum}[(i)]
\item drivers’ reaction times, 
\item subjective takeover qualities, and 
\item objective takeover qualities.\end{inparaenum} With this framework, we conduct a comprehensive analysis of the effects of drivers’ SA and SC on various aspects of takeover performance, which is the main contribution of this study. This study can provide valuable insights to readers who are interested in drivers' cognition activities during control transitions and their implications for enhancing human-vehicle interactions for safe and comfortable conditionally automated driving.





The remainder of this paper is structured as follows: Section~\ref{sec: method} details the conducted driving simulator experiment, the questionnaire for measuring drivers' cognition and subjective evaluations, the collected data, and the takeover performance prediction models; Section~\ref{sec: results} presents the prediction results and the quantified effects of situational awareness and spare capacity on various aspects of takeover performance; Section~\ref{sec: discussions} discusses the results and deliberates on the limitations of this study; and Section~\ref{sec: conclusion and future work} summarizes the findings of this study and gives recommendations for future research. 







\section{Method}
\label{sec: method}

 This study conducted a driving simulator experiment during which each participant took over vehicle control from a Conditionally Automated Driving System (CADS) across nine distinct scenarios, as detailed in Section~\ref{subsec: driving simulator experiment}. Participants provided information on their driving-related characteristics and subjective takeover experiences using the questionnaires in Section~\ref{subsec: questionnaires}. On this basis, the impacts of Situation Awareness (SA) and Spare Capacity (SC) on drivers' takeover performances (encompassing reaction times and takeover qualities) are investigated by the method in Section~\ref{subsec: data analysis}. This study is approved by the Human Research Ethics Committee (HREC) of Delft University of Technology (Case ID: 3499).

\subsection{Driving Simulator Experiment}
\label{subsec: driving simulator experiment}

Participants were recruited using online platforms (emails and LinkedIn) and offline methods (flyers), with eligibility criteria requiring a valid driver’s license and the ability to drive without glasses. An informed consent form was used to communicate the study’s objectives, procedures, duration, data anonymization protocols, and compensation details, and it was signed by all participants.

\begin{figure}[ht]
    \centering
    \begin{subfigure}{0.45\textwidth}
        \centering
        \includegraphics[width=\textwidth]{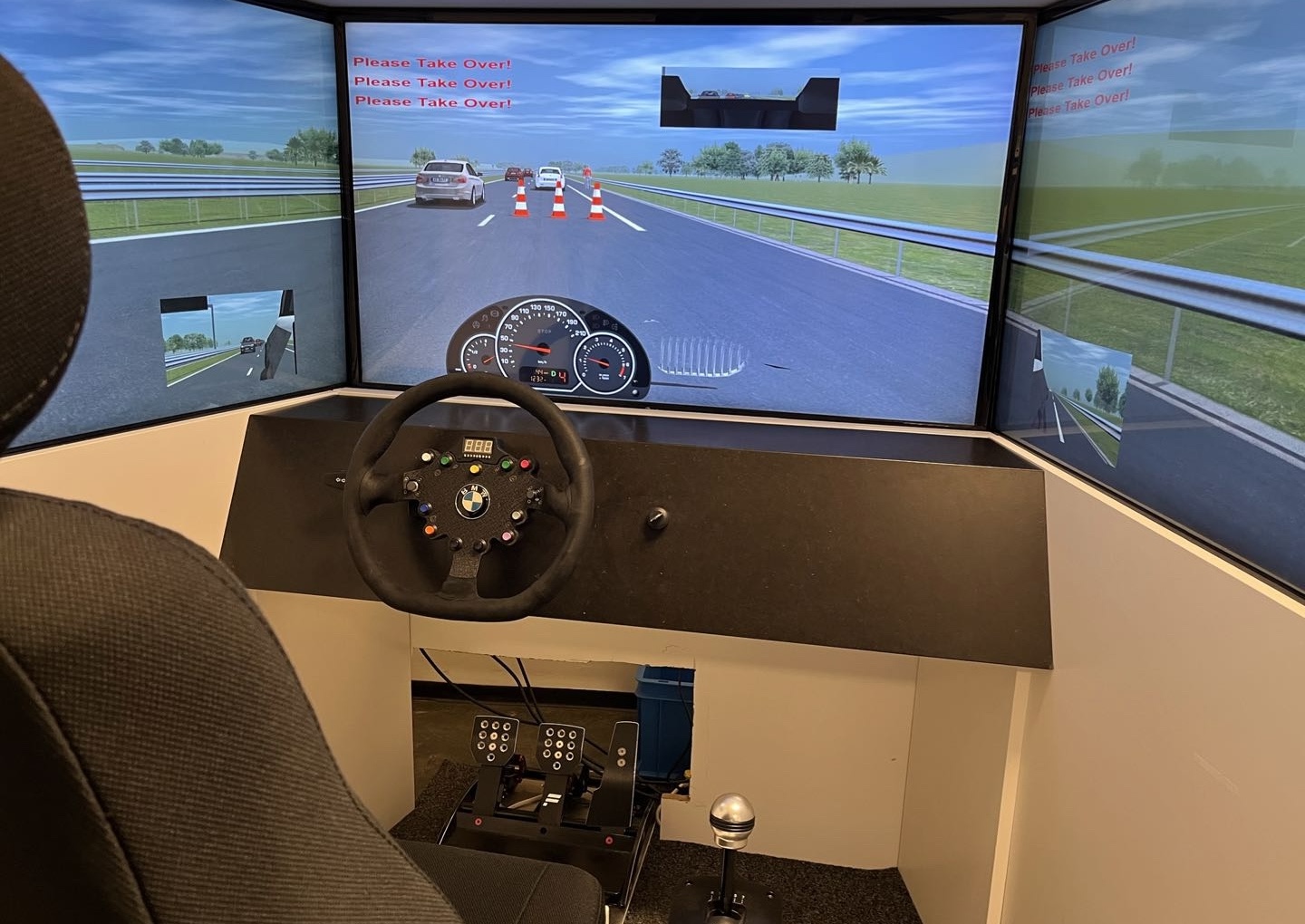} 
        \caption{Driving simulator at TU Delft.}
        \label{subfig: simulator} 
    \end{subfigure}
    \hfill
    \begin{subfigure}{0.45\textwidth}
        \centering
        \includegraphics[width=\textwidth]{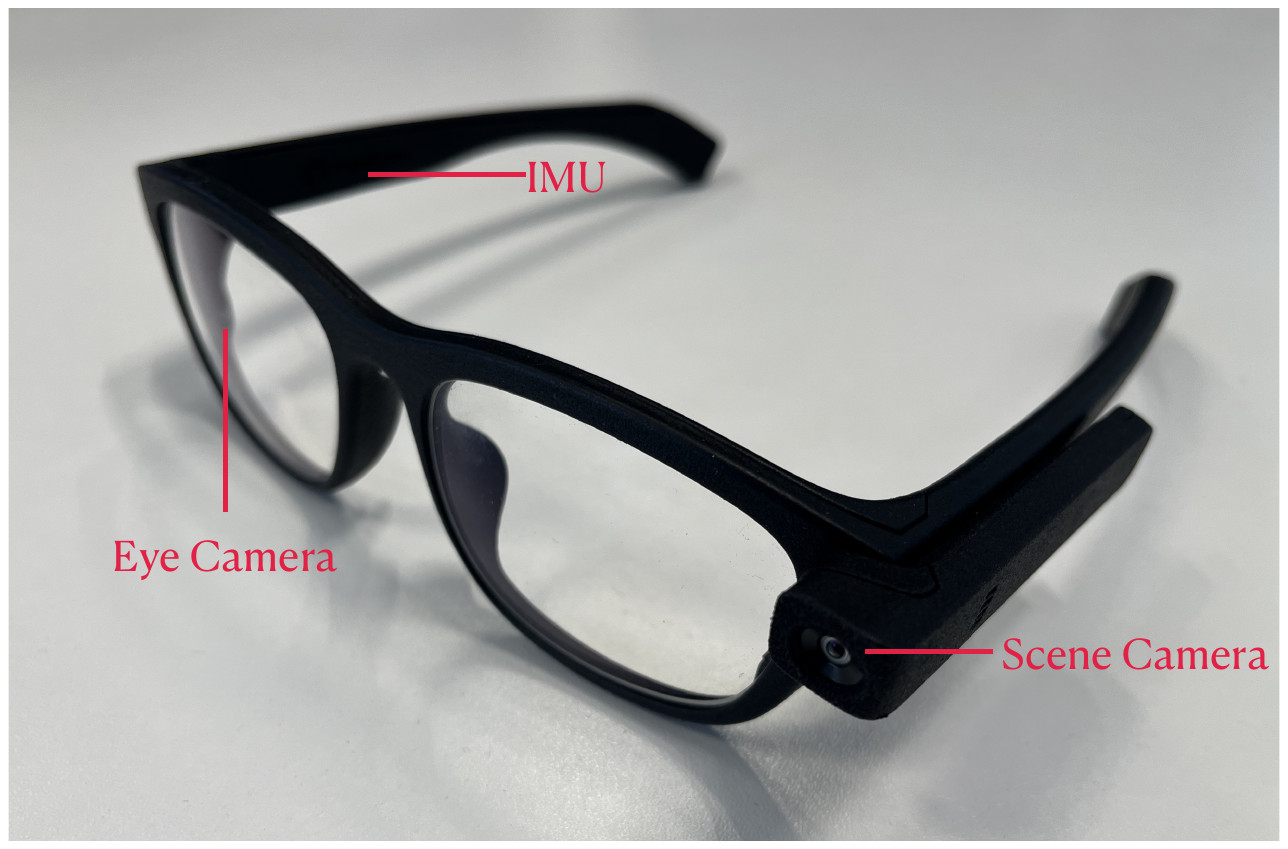} 
        \caption{Pupil Invisible eye tracking glasses}
        \label{subfig: glasses} 
    \end{subfigure}
    \caption{Instrumentation in the study.}
    \label{fig:overall} 
\end{figure}

The experiment was conducted using a fixed-base, medium-fidelity driving simulator at Delft University of Technology. As shown in Figure~\ref{subfig: simulator}, the simulator features a driver’s seat, steering wheel, gas and brake pedals, and a dashboard, with three 4K screens providing a panoramic view through the windshield and side windows. To record drivers' eye movements, this study uses a deep-learning-powered Pupil Invisible eye tracking system from Pupil Labs \citep{tonsen2020high}. The Pupil Invisible glasses, as shown in Figure~\ref{subfig: glasses}, are equipped with 
\begin{inparaenum}[(i)]
    \item a binocular pair of infrared cameras capturing eye movements at $200$ Hz with a resolution of $192*192$ px,
    \item a scene camera recording environment videos at 30 Hz with a resolution of $1088*1080$ px, and
    \item an Inertial Measurement Unit (IMU) capturing the rotation speed and translational acceleration of the glasses.
\end{inparaenum}

The experiment comprises nine takeover scenarios, varying in traffic densities and Non-Driving Related Tasks (NDRTs). Traffic densities are manipulated by introducing 0, 10, or 20 vehicles per kilometre, while NDRTs are induced using $n$-back tasks \footnote{In the $n$-back task, participants observe a sequence of positions of a blue box and are instructed to press a button when the new position is the same as the one that occurs $n$ steps back in the sequence \citep{liang2024examining}.} where $n = 0, 1, 2$. These scenarios are arranged using a Latin Square design to ensure balanced exposure \citep{calvert2014application}. All takeover scenarios occur on a two-lane motorway with a 100 km/h speed limit. The CADS allows participants to engage in NDRTs while in automated driving mode. However, a takeover request is triggered seven seconds before the ego vehicle would collide with a collision ahead which marks the CADS’s operational boundary. This request is communicated through three audible beeps and text alerts displayed in the top-left corner of the windshield. Participants are instructed to immediately resume vehicle control from CADS upon receiving the request. To reduce simulator sickness, takeover requests are only initiated on straight roads.

The experiment involves three procedures: \begin{itemize}
    \item [\textbf{(1)}] \textbf{Preparation procedure:} Participants are briefed on the CADS and the $n$-back task, followed by a practice drive to familiarize them with the simulator. Additional practice is allowed upon request, and participants are asked to report any discomfort.
    \item [\textbf{(2)}] \textbf{Takeover procedure:} Each participant completes a total of nine takeover sessions. Each session begins in automated driving mode, with a takeover request randomly triggered between 30 to 60 seconds after the session starts. Upon receiving the request, participants are required to take over vehicle control from the CADS, change lanes to bypass the detected collision, and then hand over control back to the CADS.
    \item [\textbf{(3)}] \textbf{Questionnaire procedure:} After each takeover session, participants complete a driver experience questionnaire to provide subjective ratings of their most recent takeover. Once all sessions are completed, participants fill out an additional questionnaire on driver characteristics. These two questionnaires are detailed in Section~\ref{subsec: questionnaires}.
\end{itemize}



\subsection{Questionnaires}
\label{subsec: questionnaires}

In this study, two questionnaires are employed to gather data on drivers' subjective experiences of takeovers and their individual characteristics that may affect their takeover performance. The questions involved are presented randomly to reduce order effects. Specifically, \begin{itemize}

    \item [\textbf{(1)}] \textbf{Driver Experience Questionnaire:} assesses drivers' subjective ratings of their takeover experiences. Driver experience in this study covers three aspects: 
    
    \begin{inparaenum}[(i)]
        \textit{\item Situational Awareness ($S\!A$):} this study measures $S\!A_{understanding}$ and $S\!A_{spare\_attention}$ ($S\!A_{attention\_supply}$ - $S\!A_{attention\_demand}$) using questions from the three-dimensional version of Situation Awareness Rating Technique (SART) \citep{taylor2017situational}. Additionally, two other aspects of $S\!A$ are assessed based on SART, namely $S\!A_{arousal}$ (i.e., the driver's level of alertness prior to receiving a takeover request) and $S\!A_{projection}$ (i.e., the driver's ability to anticipate the steps needed to safely regain vehicle control);

        \textit{\item Spare Capacity ($S\!C$):} \cite{fuller2011driver} defined $S\!C$ as the difference between a driver's task capability ($T\!C$) and the task demand ($T\!D$), grounded in TCI theory. In this study, $T\!C$ and $T\!D$ are measured using questions adapted from established literature \citep{zhang2019determinants, rosenbloom2010parental, pauzie2008method, lajunen1995driving, hart1988development}, tailored for takeover contexts. A full description of the questionnaire is available in our previous work \citep{liang2024examining};
        
        \textit{\item Subjective Takeover Qualities ($subjQ$):}
        This study gathers drivers' subjective evaluations of their performance for each takeover across three dimensions, namely time sufficiency ($subjQ_{suf\!f\!iciency}$), perceived risk ($subjQ_{risk}$), and performance satisfaction ($subjQ_{satis\!f\!action}$). Specifically, $subjQ_{suf\!f\!iciency}$ is measured using a reverse-phrased question, where participants are asked: ``To complete the required bypass maneuvers safely and comfortably, how many seconds would you like to adjust the time that you were provided with to take over car control? Please use this sliding scale to indicate the amount of time you want to adjust, from -10 (decrease by 10 seconds) to +10 (increase by 10 seconds)''. A higher value reflects a perceived insufficiency in the time provided. All responses are normalized to a scale of [-1, 1]. Regarding $subjQ_{risk}$, participants are asked to respond to the statement, ``I was worried about being involved in a traffic accident during the takeover process,'' indicating their agreement on a five-point Likert scale (1 = Strongly Disagree, 5 = Strongly Agree). Finally, $subjQ_{satis\!f\!action}$ is measured by having participants rate their agreement with the statement: ``I was satisfied with my performance in taking over car control,'' also using a five-point Likert scale (1 = Strongly Disagree, 5 = Strongly Agree). 

    \end{inparaenum}

    \item [\textbf{(2)}] \textbf{Driver Characteristics Questionnaire:} assess drivers' characteristics that potentially affect their responses to takeover requests across different scenarios and their corresponding takeover performances. The collected driver characteristics cover the following three aspects: \begin{inparaenum}[(i)]
    
        \textit{\item demographic information:} $age$ and $gender$;

        \textit{\item skill-related factors:} accumulated driving years ($accu\!\_years$), accumulated driving distance ($accu\!\_dis$), driving frequency ($driving\!\_f\!re$), driving skill ($driving\!\_skill$), and takeover skill ($takeover\!\_skill$); and 
        
        \textit{\item style-related factors:} driver assistance usage frequency ($assist\!\_f\!re$), risk-taking attitude ($RT\!A$), trust in CADS ($trust$), reckless and careless takeover style ($st\!yle\!\_reckless$), anxious takeover style ($st\!yle\!\_anxious$), angry and hostile takeover style ($st\!yle\!\_angry$), and patient and careful takeover style ($st\!yle\!\_patient$). 
    \end{inparaenum}
    
The utilized questionnaire is developed based on well-established instruments \citep{nordhoff2023driver, nordhoff2021perceived, lu2017much, ma2010safety, taubman2004multidimensional, lajunen1995driving}. A detailed version of this Driver Characteristics Questionnaire, including all questions, can be found in the Appendix \ref{appendix a: driver characteristic questionnaire}.

\end{itemize}

An overview of the variables collected from these two questionnaires is provided in Section~\ref{subsec: data analysis}. These variables offer valuable insights into individual differences and driver cognitions during takeovers, forming the foundation for understanding how these factors affect takeover performance.

\subsection{Data Acquisition and Analysis}
\label{subsec: data analysis}

\subsubsection{Data Acquisition}
\label{subsubsec: data acquasition}

A total of 57 drivers participated in this study. Their characteristics are summarized using descriptive statistics in Table~\ref{tab: driver characteristics} in Appendix \ref{appendix b: driver distribution}. The data reveals a diverse participant pool encompassing a wide range of demographic backgrounds, driving skills, and driving styles. This diversity allows for a detailed analysis of the factors that may influence takeover performance. In this study, the XGBoost models with those driver characteristics as inputs are considered baseline models because they serve as a reference point for evaluating the contribution of additional factors, such as Situation Awareness (SA) and Spare Capacity (SC), in affecting various components of takeover performance. 


The experiment initially produces data from 513 takeovers. After excluding 16 takeovers where participants either took vehicle control before the takeover request or forgot to press the mode switch button, and removing 31 takeovers due to incomplete questionnaires and hardware malfunctions, the final dataset for analysis consists of 466 takeovers.

To understand drivers' operational, visual, and perceptual patterns for taking over vehicle control, this study records data from the 30 seconds preceding the initiation of takeover requests to the successful completion of lane changes in distinct takeover scenarios. The collected data is classified into three categories:
\begin{itemize}

    \item[(1)] \textbf{operational data}, which includes vehicle positions, driving mode, accelerations, decelerations, steering wheel angles, accelerator pedal positions, braking pedal positions, and timestamps from the driving simulator. On this basis, we calculate five operational metrics, namely $t_{buttion}$, $t_{steering}$, $t_{pedal}$, $T\!oT$, and $TTC$, as detailed in Table~\ref{tab: operational metric} in Appendeix \ref{appendix c: metric collection}.

    \item[(2)] \textbf{visual data}, which includes drivers' gaze position, fixation position \footnote{Samples with gaze velocities below 900 px/s are classified as fixations \citep{drews2024strategies}.}, and the corresponding timestamps from Pupil Invisible eye tracking glasses. On this basis, 13 visual metrics are extracted as indicators of drivers' SA, as outlined in Table~\ref{tab: sa visual metric definition} in Appendeix \ref{appendix c: metric collection}. Another 12 visual metrics are extracted as indicators of drivers' SC, as detailed in Table~\ref{tab: sc visual metric definition} in the same Appendeix. These metrics provide objective measures of SA and SC.
    

    \item[(3)] \textbf{self-reported data}, which includes drivers' perception of their characteristics, SA, SC, and takeover experience across various takeover scenarios. These data are collected using the questionnaires detailed in Section~\ref{subsec: questionnaires}, thus presenting subjective measurements of the relevant variables.
\end{itemize}

To provide an overview of the variables investigated, we summarize the factors that may influence driver takeover performance in Table~\ref{tab: potential variables}, which captures a range of individual characteristics and driver cognition of takeover tasks. Meanwhile, Table~\ref{tab: performance metrics} outlines the key metrics used to evaluate takeover performance in terms of reaction times, subjective takeover qualities, and objective takeover qualities. These metrics serve as critical indicators for assessing how well drivers handle the takeover process from efficiency, safety, and comfort perspectives.

\begin{table*}[!h]
\centering
\caption{Overview of investigated factors that potentially affect takeover performance.}
\label{tab: potential variables}
\resizebox{\textwidth}{!}{ 
\begin{tabular}{p{6em} l p{32em}}

\toprule
\textbf{category} & \textbf{sub-category} &\textbf{variables} \\ \midrule
\multirow{3}{8em}{Driver Characteristics ($D\!C$)} & demographics & $age$, $gender$\\
& skill-related factors & $accu\!\_years$, $accu\!\_dis$, $driving\!\_f\!re$, $driving\!\_skill$, $takeover\!\_skill$, $assist\!\_f\!re$\\
& style-related factors & $RT\!A$, $trust$, $st\!yle\!\_reckless$, $st\!yle\!\_anxious$, $st\!yle\!\_angry$, $st\!yle\!\_patient$ \\
                                           \midrule

\multirow{3}{6em}{Situational Awareness ($S\!A$)} & subjective measurements &$S\!A_{understanding}$, $S\!A_{spare\_attetion}$, $S\!A_{arousal}$, $S\!A_{projection}$ \\

& objective measurements  & $nr_{road}$, $nr_{rearview}$, $nr_{sideview}$, $nr_{centre}$, $nr_{dashboard}$, $f_{max}$, $f_{mean}$, $f_{std}$, $num_{f}$, $s_{max}$, $s_{mean}$, $s_{std}$, $num_{s}$\\
                                 \midrule
\multirow{4}{8em}{Spare Capacity ($S\!C$)} & subjective measurements & $T\!C$ (including $T\!C_{anticipation}$, $T\!C_{reaction}$, $T\!C_{speed\!\_adjust}$, $T\!C_{lane\_change}$, and $T\!C_{saf\!ety}$), $T\!D$ (including $T\!D_{mental}$, $T\!D_{visual}$, and $T\!D_{temporal}$) \\
& objective measurements & $t_{road}$, $t_{HMI}$, $t_{mirror}$, $PCT_{road}$, $PCT_{HMI}$, $PCT_{mirror}$, $N\!O_{road}$, $N\!O_{HMI}$, $N\!O_{mirror}$, $AV\!G_{road}$, $AV\!G_{HMI}$, $AV\!G_{mirror}$ \\
                                 \bottomrule
\end{tabular}}
\end{table*}

\begin{table*}[!h]
\centering
\caption{Overview of metrics for takeover performance (TOR: Takeover Request).}
\label{tab: performance metrics}
\resizebox{\textwidth}{!}{ 
\begin{tabular}{p{7.5em} l l l}

\toprule
\textbf{category} &  \textbf{metric} &\textbf{unit} & \textbf{description} \\ \midrule

\multirow{3}{7.5em}{reaction times}&$t_{button}$ & $s$  & time to press the mode switch button to enable manual inputs following a TOR\\
                                           & $t_{road}$ & $s$ & time to first visually fixate on the road following a TOR \\
                                           & $T\!oT$ & $s$ & time to perform the first conscious operational response to a TOR\\
                                           \midrule

\multirow{3}{7.5em}{subjective takeover qualities} & $subjQ_{suf\!f\!iciency}$ & -- & perceived sufficiency of time provided for a takeover\\
                                 & $subjQ_{risk}$ & -- & perceived risk of an accident during a takeover\\
                                 & $subjQ_{satis\!f\!action}$ & -- & perceived satisfaction for their performance during a takeover\\
                                 \midrule

\multirow{4}{7.5em}{objective takeover qualities} & $objQ_{ttc}$ & $s$ & minimum time to collision from TOR to lane change\\
                                 & $objQ_{steer}$ & $rad$ & maximum steering wheel angle from TOR to lane change\\
                                 & $objQ_{acc}$ & $m/s^2$ & maximum acceleration from TOR to lane change\\
                                 & $objQ_{dec}$ & $m/s^2$ & maximum deceleration from TOR to lane change\\
                                 \bottomrule
\end{tabular}}

\end{table*}


\subsubsection{Data Analysis}
\label{subsubsec: data analysis}

Based on the collected data, this study employs the eXtreme Gradient Boosting (XGBoost) model \citep{chen2016xgboost} to explore the effects of drivers' Situational Awareness (SA) and Spare Capacity (SC) on diverse aspects of takeover performance. XGBoost has achieved widespread recognition for its success in numerous machine learning competitions \citep{nielsen2016tree} and has been applied in automated driving research to predict takeover time \citep{chen2024predicting,ayoub2022predicting} and takeover quality \citep{zhu2023takeover}. In this study, XGBoost models are developed to predict various takeover performance metrics using different input combinations: \begin{inparaenum}[(i)]
    \item driver characteristics ($D\!C$);
    \item $D\!C$ combined with Situational Awareness ($S\!A$);
    \item $D\!C$ combined with Spare Capacity ($S\!C$); and
    \item $D\!C$ combined with both $S\!A$ and $S\!C$.
\end{inparaenum} 
The prediction results are evaluated using Root Mean Squared Error (RMSE) and Mean Absolute Error (MAE) \citep{antypas2024time, ayoub2022predicting, yang2021personalized}. By comparing the results across XGBoost models with different input combinations, this study examines the relative contributions of $D\!C$, $S\!A$, and $S\!C$ in explaining variations in each performance metric.

Furthermore, this study delves into the contributions of individual predictors to variations in takeover performance metrics. Bonferroni-adjusted significance tests are conducted to identify key predictors from the identified effective input combinations. These significant features are then incorporated into XGBoost models to interpret their impact on performance metrics. Using 10-fold cross-validation repeated 100 times, the models compute the average importance of each input feature to ensure the stability and robustness of the results. Additionally, SHapley Additive exPlanation (SHAP) values \citep{lundberg2017unified} are employed to quantify both the direction and magnitude of each predictor’s influence. Note that the order of features when sorted by SHAP values may differ from the order sorted based on feature importance. This is because \begin{inparaenum}[(i)]
    \item SHAP values account for interactions among features, while feature importance only considers the contribution of individual features in isolation; and,
    \item feature importance measures how much a feature contributes to reducing loss, while SHAP values measure both magnitude and direction of a feature's impact.
\end{inparaenum}
As a result, some features may have high split importance in XGBoost but low overall impact in SHAP rankings. This study employs both methods: feature importance for high-level insights and feature selection, and SHAP values for detailed interpretability and a deeper understanding of feature contributions.

In Section \ref{sec: discussions}, the performance of XGBoost models is compared with two additional machine learning models, namely Random Forest (RF) and Light Gradient Boosting Machine (LightGBM), to strengthen the internal validity of this study. This comparison ensures that the findings are robust and not specific to a single modeling approach. Furthermore, the results of this study are benchmarked against existing literature to establish external validity, demonstrating their alignment with or divergence from prior research in the field.


\section{Results}
\label{sec: results}

This section presents the effects of Situational Awareness (SA) and Spare Capacity (SC) on ten takeover performance metrics which are grouped into three categories: reaction times (Section \ref{subsec: reaction times}), subjective takeover quality (Section \ref{subsec: subjective quality}), and objective takeover quality (Section \ref{subsec: objective quality}). For each metric, we report the performance of XGBoost models using different feature sets, identify significant predictors through Bonferroni-adjusted tests, and interpret feature contributions using SHAP values.


\subsection{Reaction Times}
\label{subsec: reaction times}

Three temporal metrics are selected for quantifying distinct phases of drivers' reaction processes following takeover requests, namely \begin{inparaenum}[(i)]
    \item button press time ($t_{button}$) -- the time taken to press the mode switch button, reflecting drivers' behavioral reaction speed;
    \item road reorientation time ($t_{road}$) -- the time taken to shift gaze back to the forward roadway, indicating drivers' visual reorientation speed; and
    \item takeover time ($T\!oT$) -- the time to consciously take over control of the vehicle, capturing drivers' cognitive response speed.
\end{inparaenum}

\subsubsection{Button press time}

The performance evaluations for $t_{button}$ models are summarized in Table~\ref{tab: t_button}. The baseline model using only driver characteristics (XGBTbutton$_{dc}$) achieves reasonable performance but records the highest RMSE and MAE among the four models. Incorporating SA (XGBTbutton$_{dc+sa}$) significantly enhances model performance, reducing RMSE and MAE by 9.21\% and 9.91\% respectively ($p_{adjusted} < 0.01$). Adding SC (XGBTbutton$_{dc+sc}$) leads to minor improvements compared with XGBTbutton${dc}$, with slight reductions in both RMSE (2.79\%) and MAE (3.18\%). However, these changes are statistically insignificant (both $p_{adjusted} > 0.05$). Combining both SA and SC (XGBTbutton$_{dc+sa+sc}$) results in the most accurate representation of $t_{button}$, with RMSE decreasing by 10.48\% and MAE by 10.59\% (both $p_{adjusted} < 0.01$) compared to XGBTbutton$_{dc}$. Furthermore, XGBTbutton$_{dc+sa+sc}$ significantly decreases RMSE by 7.92\% and MAE by 7.65\% compared to XGBTbutton$_{dc+sc}$ ($p_{adjusted} < 0.05$), but shows no significant difference when compared to XGBTbutton$_{dc+sa}$ ($p_{adjusted} = 1.00$). These results suggest that SA plays a more dominant role than SC in explaining the variations in $t_{button}$.

\begin{table}[htbp]
\centering
\caption{Performance of XGBoost-based $t_{button}$ models.}
\label{tab: t_button}
\begin{tabular}{llcc}
\toprule
\textbf{Model}& \textbf{Inputs} & \textbf{RMSE ($\downarrow$)} & \textbf{MAE ($\downarrow$)}  \\ \midrule
XGBTbutton$_{dc}$& $D\!C$  & 0.6848 & 0.4966   \\ 

XGBTbutton$_{dc+sa}$ & $D\!C$ $+$ $S\!A$  & 0.6217 & 0.4474  \\ 

XGBTbutton$_{dc+sc}$ & $D\!C$ $+$ $S\!C$ & 0.6657 &	0.4808  \\ 

XGBTbutton$_{dc+sa+sc}$ & $D\!C$ $+$ $S\!A$ $+$ $S\!C$  & 0.6130 & 0.4440  \\ 
\bottomrule
\end{tabular}
\end{table}

\begin{figure*}[!h]
    \centering
    \includegraphics[width=0.7\columnwidth]{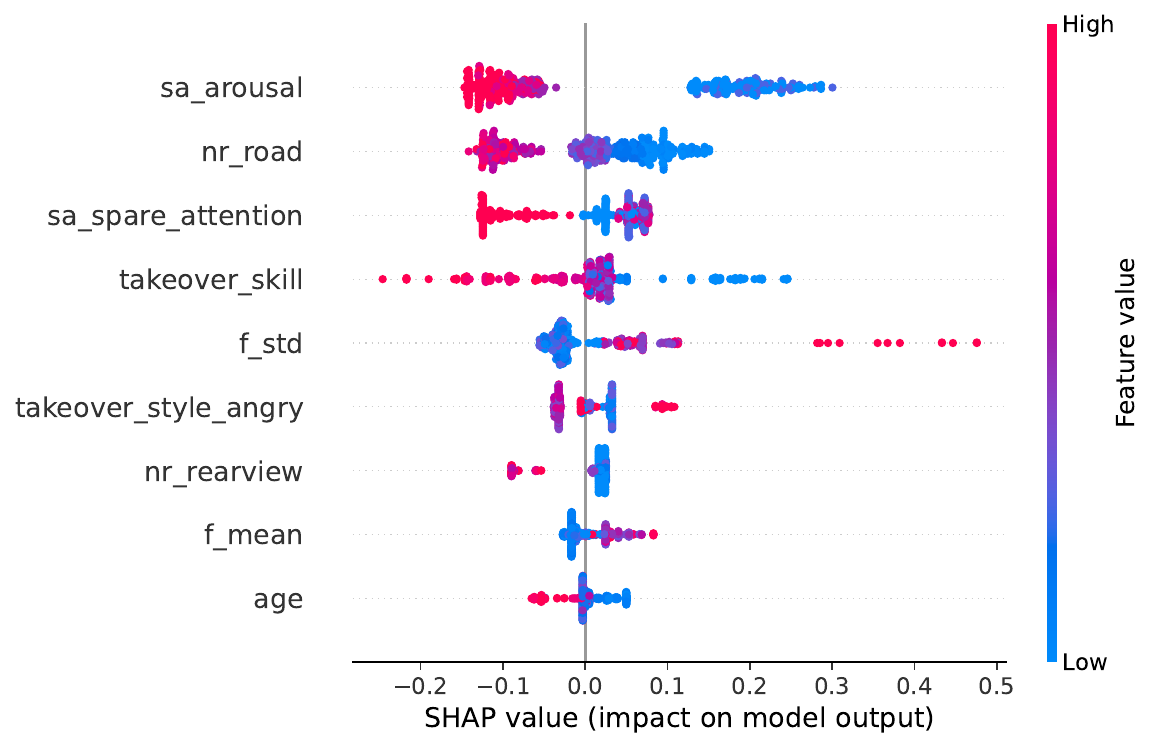}
    \caption{Summary plot of significant Driver Characteristics ($D\!C$) and Situational Awareness ($S\!A$) factors influencing the button press time ($t_{button}$).}
    \label{fig: t_button summary plot}
\end{figure*}

Building on the significant impacts of DC and SA on $t_{button}$, this study further explores the contributions of individual factors. Bonferroni tests identify nine significant factors that significantly influence $t_{button}$ ($p_{adjusted} < 0.01$), which are then incorporated into the XGBoost model. Their feature importance and Bonferroni test results ($p_{adjusted}$) are summarized in Table \ref{tab: t_button significant factors}. The results indicate that SA-related factors play a dominant role, whereas DC-related factors have a smaller yet significant influence. Among all nine factors, $S\!A\_{arousal}$ (the driver’s level of alertness before receiving a takeover request) emerges as the most critical factor affecting $t_{button}$, followed by two additional SA-related factors: $nr_{road}$ and $S\!A_{spare_attention}$. Besides, the summary plot in Figure~\ref{fig: t_button summary plot} illustrates that, generally, drivers with higher SA (as indicated by higher arousal levels before the takeover request, greater spare attention, more frequent glances at driving-related AOIs such as the road and rearview mirror, and shorter, more stable visual fixation durations) tend to react and press the mode switch button more quickly.

\begin{table}[htbp]
\centering
\caption{Feature importance and Bonferroni test results ($p_{adjusted}$) for significant factors of $t_{button}$.}
\label{tab: t_button significant factors}
\begin{tabular}{lcc l l c c}
\toprule
\textbf{Feature} & \textbf{Importance ($\uparrow$)} & \textbf{$p_{adjusted}$ ($\downarrow$)} & &\textbf{Feature} & \textbf{Importance ($\uparrow$)} & \textbf{$p_{adjusted}$ ($\downarrow$)} \\ \midrule

$S\!A_{arousal}$ & 0.2739 & 1.2234e-43 & & $f_{std}$ & 0.0758 & 1.6861e-19 \\ 
$nr_{road}$ & 0.1565 & 3.4159e-23 & & $nr_{rearview}$ & 0.0643 & 3.6289e-06\\
$S\!A_{spare\_attention}$ & 0.1225 & 2.5558e-21 & & $takeover\!\_skill$  & 0.0635 & 7.0540e-15\\
$st\!yle\!\_angry$  & 0.1043 & 1.5503e-15 & &$age$ & 0.0441 & 7.0513e-05 \\ 
$f_{mean}$ & 0.0951 & 3.3310e-04 & & & &\\

\bottomrule
\end{tabular}
\end{table}


\subsubsection{Road reorientation time}

Table~\ref{tab: t_road} presents the results of modeling road reorientation time ($t_{road}$), which represents the time drivers take to reorient their visual attention to the road following a takeover request. The baseline model (XGBTroad$_{dc}$) exhibits the highest RMSE and MAE. XGBTroad$_{dc+sa}$ incorporating SA-related factors into the baseline model significantly improves performance ($p_{adjusted} < 0.01$), reducing RMSE by 9.68\% and decreasing MAE by 12.81\%. Adding SC-related features (XGBTroad$_{dc+sc}$) to the baseline model does not yield statistically significant changes, as indicated by $p_{adjusted} = 1.00$ for both RMSE and MAE. The combined model incorporating both SA and SC features (XGBTroad$_{dc+sa+sc}$) reduces RMSE by 10.05\% and decreases MAE by 12.47\% ($p_{adjusted} < 0.01$). Notably, there is no significant difference between XGBTroad$_{dc+sa}$ and XGBTroad$_{dc+sa+sc}$ ($p_{adjusted} = 1.00$). These results indicate that SA accounts for a substantial portion of the explainable variation in $t_{road}$, and the additional contribution of SC remains insignificant.

\begin{table}[htbp]
\centering
\caption{Performance of XGBoost-based $t_{road}$ models.}
\label{tab: t_road}
\begin{tabular}{llcc}
\toprule
\textbf{Model}& \textbf{Inputs} & \textbf{RMSE ($\downarrow$)} & \textbf{MAE ($\downarrow$)}\\ \midrule
XGBTroad$_{dc}$& $D\!C$  & 0.6973 & 0.5332   \\ 

XGBTroad$_{dc+sa}$ & $D\!C$ $+$ $S\!A$  & 0.6298 & 0.4649  \\ 

XGBTroad$_{dc+sc}$ & $D\!C$ $+$ $S\!C$ & 0.6889 & 0.5400  \\ 

XGBTroad$_{dc+sa+sc}$ & $D\!C$ $+$ $S\!A$ $+$ $S\!C$  & 0.6273 &	0.4667  \\ 
\bottomrule
\end{tabular}
\end{table}

\begin{figure*}[!h]
    \centering
    \includegraphics[width=0.7\columnwidth]{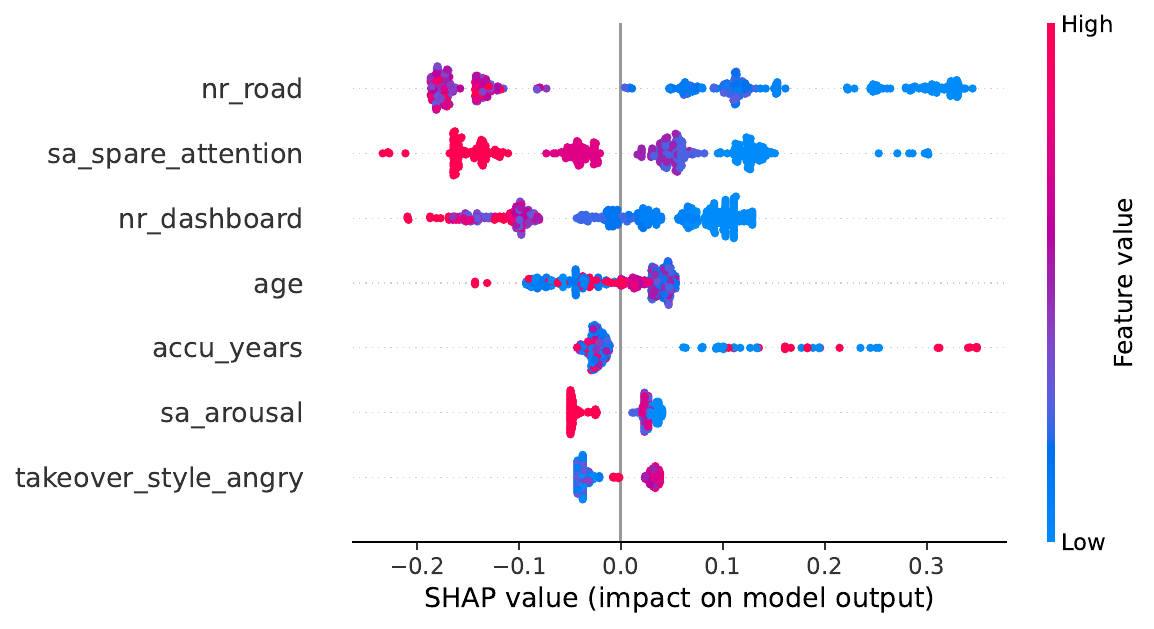}
    \caption{Summary plot of significant Driver Characteristics ($D\!C$) and Situational Awareness ($S\!A$) factors influencing the road reorientation time ($t_{road}$).}
    \label{fig: t_road summary plot}
\end{figure*}

 Seven significant factors are identified via Bonferroni tests ($p_{adjusted} < 0.01$) and incorporate them into the XGBoost model to interpret their contributions to $t_{road}$. As shown in Table~\ref{tab: t_road significant factors}, SA-related factors play a dominant role, with objective SA factors being more influential than subjective ones. In particular, $nr_{road}$ stands out as the most influential factor, likely because it directly reflects drivers' visual attention allocation strategies. Besides, the summary plot in Figure~\ref{fig: t_road summary plot} shows that drivers with higher SA (i.e., those who frequently glance at the road and dashboard, and perceive themselves as having greater spare attention and higher arousal levels) generally spend less time returning their gaze to the road after a takeover request.

\begin{table}[htbp]
\centering
\caption{Feature importance and Bonferroni test results ($p_{adjusted}$) of significant factors of $t_{road}$.}
\label{tab: t_road significant factors}
\begin{tabular}{lcc l l c c}
\toprule
\textbf{Feature} & \textbf{Importance ($\uparrow$)} & \textbf{$p_{adjusted}$ ($\downarrow$)} & &\textbf{Feature} & \textbf{Importance ($\uparrow$)} & \textbf{$p_{adjusted}$ ($\downarrow$)} \\ \midrule

$nr_{road}$ & 0.4588 & 1.8133e-43  & & $accu\!\_years$ & 0.0725 & 9.8633e-15 \\
$nr_{dashboard}$ & 0.1230 & 3.2160e-23 & & $age$ & 0.0691 & 5.2440e-11 \\
$S\!A_{spare\_attention}$ & 0.1137 & 9.9628e-22 & & $st\!yle\!\_angry$ & 0.0690 & 1.1174e-04 \\
$S\!A_{arousal}$ & 0.0940 & 5.9896e-06 & &  & & \\

\bottomrule
\end{tabular}
\end{table}


\subsubsection{Takeover time}

The performance evaluations of XGBoost-based $T\!oT$ models are summarized in Table~\ref{tab: tot}. Modeling $T\!oT$ is more challenging compared with $t_{button}$ and $t_{road}$, as indicated by the relatively higher RMSE and MAE values across all models. Incorporating SA and SC (separately or jointly) significantly improves model accuracy over using DC alone ($p_{adjusted} < 0.05$). Specifically, compared to the baseline model (XGBToT$_{dc}$), incorporating only SA (XGBToT$_{dc+sa}$) reduces RMSE by 5.61\% and MAE by 5.52 \%. Adding SC alone (XGBToT$_{dc+sc}$) shows a greater improvement over the baseline than SA alone, with RMSE and MAE reductions of 11.80\% and 10.55\%, respectively. Combining both SA and SC (XGBToT$_{dc+sa+sc}$) provides the best performance, with reductions of 14.01\% in RMSE and 12.55\% in MAE, compared to the baseline. These findings highlight that while both SA and SC contribute meaningfully to the model’s ability to explain $T\!oT$, SC appears to have a more substantial impact. Besides, Bonferroni-corrected tests reveal no significant improvement of XGBToT$_{dc+sa+sc}$ over XGBToT$_{dc+sc}$ ($p_{adjusted} > 0.05$), indicating that the impact of SA becomes negligible when SC is already accounted for.


\begin{table}[htbp]
\centering
\caption{Performance of XGBoost-based $T\!oT$ models.}
\label{tab: tot}
\begin{tabular}{llcc}
\toprule
\textbf{Model}& \textbf{Inputs} & \textbf{RMSE ($\downarrow$)} & \textbf{MAE ($\downarrow$)}  \\ \midrule
XGBToT$_{dc}$& $D\!C$  & 1.3082 &	1.0068  \\ 

XGBToT$_{dc+sa}$ & $D\!C$ $+$ $S\!A$  & 1.2348 & 0.9512  \\ 

XGBToT$_{dc+sc}$ & $D\!C$ $+$ $S\!C$ & 1.1538 &	0.9006 \\ 

XGBToT$_{dc+sa+sc}$ & $D\!C$ $+$ $S\!A$ $+$ $S\!C$  & 1.1249 & 0.8804\\ 
\bottomrule
\end{tabular}
\end{table}

This study examines the importance of individual DC- and SC-related factors in modeling $T\!oT$. Bonferroni tests identify ten significant factors for $T\!oT$ ($p_{adjusted} < 0.05$) and all of them are related to SC. This underscores the critical role of SC in explaining variability in $T\!oT$, while more stable individual differences among drivers (e.g., driving experience and style) may be less impactful on $T\!oT$ when SC-related factors are already considered. Average feature importance and Bonferroni test results ($p_{adjusted}$) of these factors are summarized in Table \ref{tab: tot significant factors}. Overall, subjective SC-related factors show greater impacts on $T\!oT$ than objective SC-related gaze metrics with higher feature importance. Particularly, $SC_{TC}$ emerges as the most influential feature, suggesting that drivers' self-assessed task management ability can more accurately explain the time required for drivers to return to the cognitive driving loop.

\begin{table}[htbp]
\centering
\caption{Feature importance and Bonferroni test results ($p_{adjusted}$) of significant factors of $T\!oT$.}
\label{tab: tot significant factors}
\begin{tabular}{lcc l l c c}
\toprule
\textbf{Feature} & \textbf{Importance ($\uparrow$)} & \textbf{$p_{adjusted}$ ($\downarrow$)} & &\textbf{Feature} & \textbf{Importance ($\uparrow$)} & \textbf{$p_{adjusted}$ ($\downarrow$)} \\ \midrule

$SC_{TC}$ & 0.2553 & 7.1117e-12 & & $DU\!R_{H\!M\!I}$ & 0.0699 & 1.4399e-16 \\
$SC$ & 0.1634 & 7.0952e-18 & & $t_{mirror}$ & 0.0688 & 5.5131e-11 \\
$SC_{T\!D}$ & 0.1079 & 5.3343e-08 & & $AV\!G_{mirror}$ & 0.0683 & 7.0202e-12 \\
$t_{H\!M\!I}$ & 0.0853 & 5.7321e-10 & & $AV\!G_{H\!M\!I}$ & 0.0591 & 3.8123e-22 \\
$N\!O_{mirror}$ & 0.0775 & 1.0248e-18 & & $AV\!G_{road}$ & 0.0446 & 6.4752e-06 \\

\bottomrule
\end{tabular}
\end{table}

\begin{figure*}[!h]
    \centering
    \includegraphics[width=0.7\columnwidth]{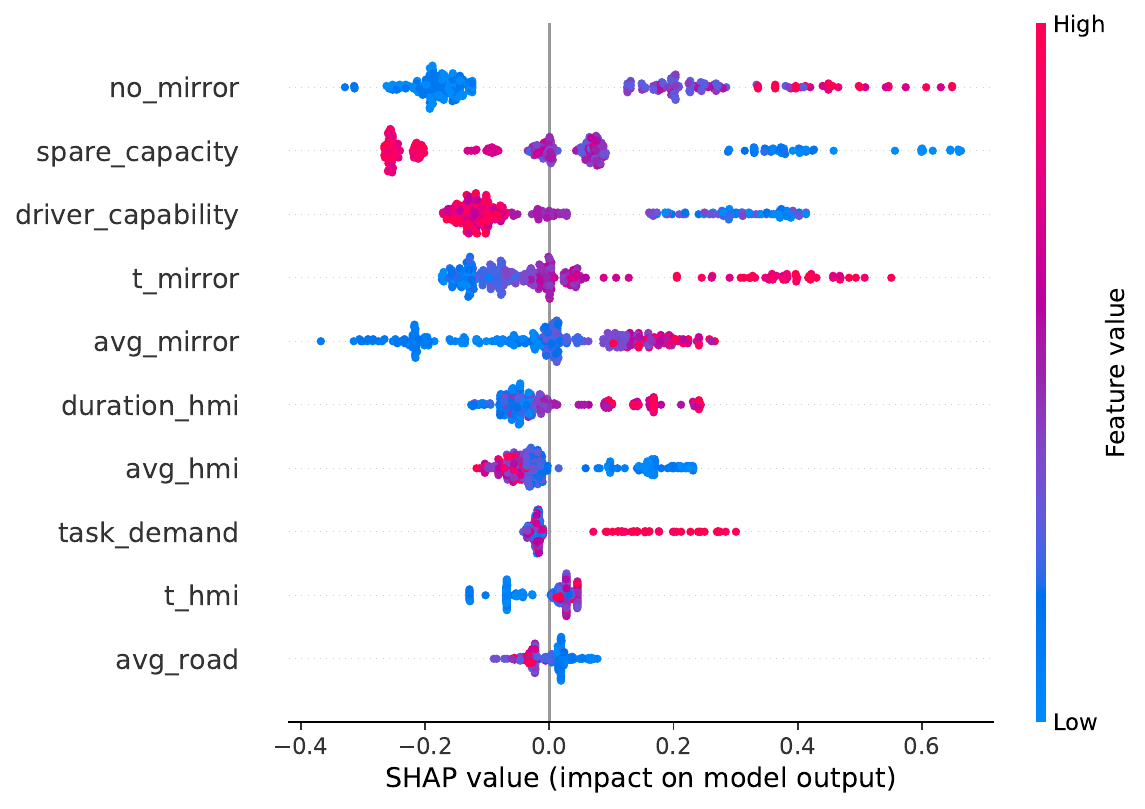}
    \caption{Summary plot of significant Spare Capacity ($S\!A$) factors influencing drivers' takeover time ($T\!oT$).}
    \label{fig: tot summary plot}
\end{figure*}

We further explore the impact of the identified ten significant SC-related factors on $T\!oT$. As illustrated in Figure~\ref{fig: tot summary plot}, drivers who perceive themselves as having lower SC (i.e., those who think the task is more demanding and their capability to handle the task is lower) generally spend more time resuming conscious vehicle control. Besides, gaze interactions with three AOIs (side mirrors, HMI, and the forward road) are important for capturing $T\!oT$, while the visual interaction patterns vary across these AOIs. Generally, drivers with longer $T\!oT$ tend to engage in more extensive side mirror checks, take longer to make their first glance at both the side mirrors and HMI, spend more total time interacting with the HMI, and exhibit longer fixation durations on mirrors and the HMI, while spending less time per fixation on the forward road. These findings highlight the critical role of both self-perceived SC and gaze indicators in understanding and modeling drivers' $T\!oT$.

\subsubsection{Summary}

Two main patterns in drivers' reaction times following takeover requests are identified: \begin{inparaenum}[(i)]
    \item Drivers' immediate responses - including mode switch activation and visual attention redirection - are primarily governed by Situational Awareness (SA), with higher SA levels corresponding to faster reactions; while
    \item SA offers limited additional explanatory power for reflective responses (such as takeover time) beyond what SC provides, with reduced SC levels consistently resulting in delayed conscious responses.
\end{inparaenum}

\subsection{Subjective Takeover Quality}
\label{subsec: subjective quality}

In this section, we examine the factors that influence drivers' perceptions of takeover quality. To capture subjective takeover quality, three key metrics are selected and measured through a questionnaire: \begin{inparaenum}[(i)]
    \item perceived sufficiency of the time provided for fulfilling the takeover task ($subjQ_{suf\!f\!iciency}$),
    \item perceived risk of being involved in an accident during the takeover ($subjQ_{risk}$), and
    \item perceived satisfaction with their performance during the takeover ($subjQ_{satis\!f\!action}$).
\end{inparaenum}


\subsubsection{Time sufficiency}

Table~\ref{tab: sq suff} presents the performance evaluation of XGBoost-based models for $subjQ_{suf\!f\!iciency}$ using different input combinations. The baseline model, XGBSQsuff$_{dc}$, which relies solely on DC as predictors, demonstrates strong performance with low RMSE and MAE. The addition of SA and SC does not significantly improve the model's explanatory power, as evidenced by the minimal changes in performance metrics and $p_{adjusted} > 0.05$. These results imply that SA and SC contribute only minimally, if at all, to explaining $subjQ_{suf\!f\!iciency}$, with their effects either overlapping with or being negligible relative to DC. 


\begin{table}[htbp]
\centering
\caption{Performance of XGBoost-based $subjQ_{suf\!f\!iciency}$ models.}
\label{tab: sq suff}
\begin{tabular}{llcc}
\toprule
\textbf{Model}& \textbf{Inputs} & \textbf{RMSE ($\downarrow$)} & \textbf{MAE ($\downarrow$)}  \\ \midrule
XGBSQsuff$_{dc}$& $D\!C$  & 0.2332 &	0.1810  \\ 

XGBSQsuff$_{dc+sa}$ & $D\!C$ $+$ $S\!A$  & 0.2325	& 0.1808  \\ 

XGBSQsuff$_{dc+sc}$ & $D\!C$ $+$ $S\!C$ & 0.2240 &	0.1712 \\ 

XGBSQsuff$_{dc+sa+sc}$ & $D\!C$ $+$ $S\!A$ $+$ $S\!C$  & 0.2232 &	0.1711\\ 
\bottomrule
\end{tabular}
\end{table}

\begin{figure*}[!h]
    \centering
    \includegraphics[width=0.7\columnwidth]{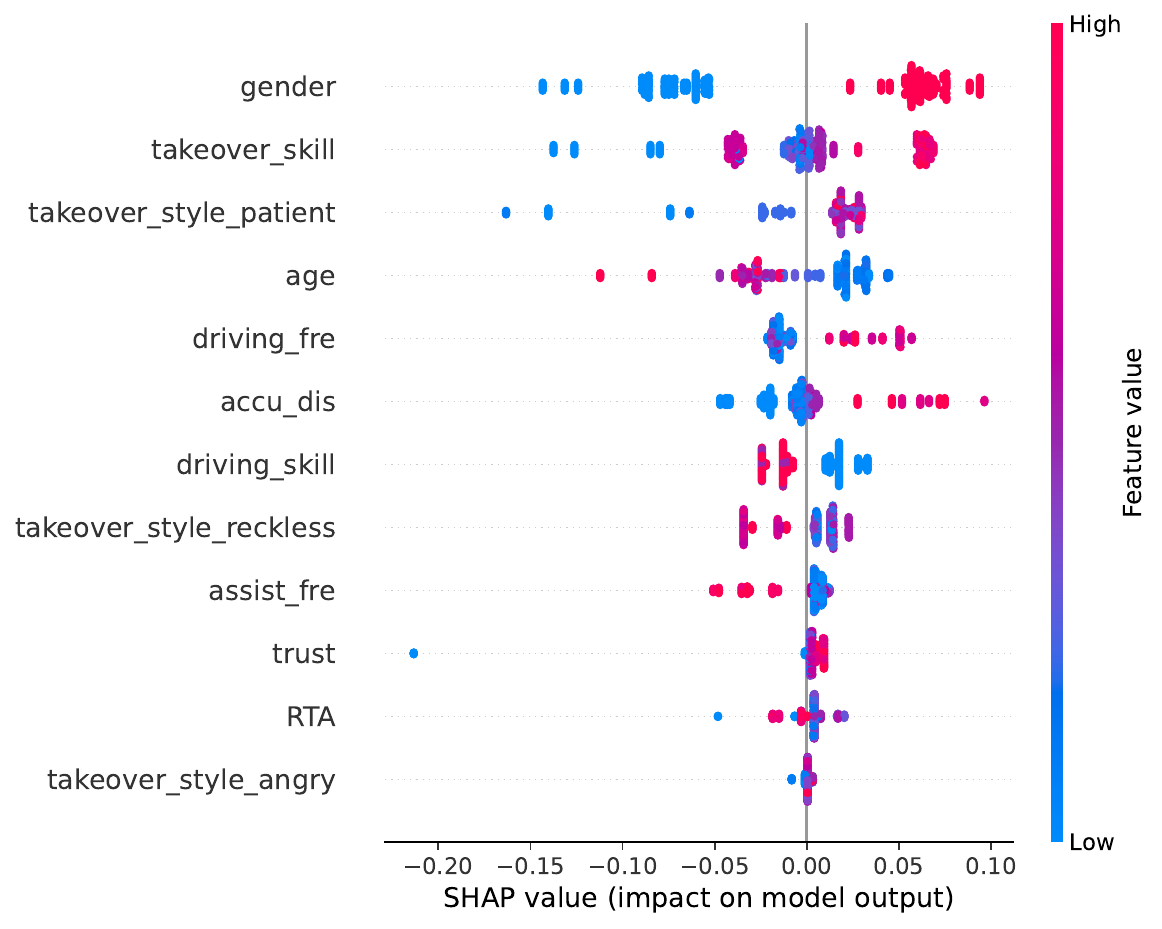}
    \caption{Summary plot of significant Driver Characteristics ($D\!C$) factors influencing drivers' perceived time sufficiency ($subjQ_{suf\!f\!iciency}$).}
    \label{fig: sufficiency summary plot}
\end{figure*}

On this basis, this study investigates the importance of specific DC-related factors in determining $subjQ_{suf\!f\!iciency}$. Bonferroni tests identify 12 significant factors ($p_{adjusted} < 0.01$), which are subsequently fed into the XGBoost-based $subjQ_{suf\!f\!iciency}$ model. Table \ref{tab: sufficiency significant factors} presents the feature importance and Bonferroni-adjusted significance ($p_{adjusted}$) of these factors. $gender$ emerges as the most influential factor in determining $subjQ_{suf\!f\!iciency}$, which may stem from a variety of socio-cultural and psychological factors that influence how individuals assess takeover conditions. Additionally, the summary plot in Figure \ref{fig: sufficiency summary plot} reveals that drivers who perceive the provided time budget for takeover tasks as less sufficient tend to be male, older, and more impulsive (e.g., less patient, more reckless, and more prone to risk-taking). They typically have lower trust in conditionally automated driving systems and diminished driving proficiency (e.g., reduced takeover skill, infrequent driving, and shorter accumulated driving distances) despite rating their driving abilities highly. Moreover, they may frequently use driver assistance systems, potentially raising their expectations for automated interventions.


\begin{table}[htbp]
\centering
\caption{Feature importance and Bonferroni test results ($p_{adjusted}$) of significant factors of $subjQ_{suf\!f\!iciency}$.}
\label{tab: sufficiency significant factors}
\begin{tabular}{lcc l l c c}
\toprule
\textbf{Feature} & \textbf{Importance ($\uparrow$)} & \textbf{$p_{adjusted}$ ($\downarrow$)} & &\textbf{Feature} & \textbf{Importance ($\uparrow$)} & \textbf{$p_{adjusted}$ ($\downarrow$)} \\ \midrule

$gender$   & 0.1820 &  9.4805e-50 & &$driving\!\_f\!re$  &  0.0733 &  1.9613e-28 \\
$st\!yle\!\_patient$    & 0.1062 &  1.2289e-41 & &$accu\!\_dis$ & 0.0681 &  8.4145e-33 \\
$driving\!\_skill$    & 0.0917 &  1.7749e-19 & &$st\!yle\!\_reckless$    & 0.0673 &  1.9803e-12 \\
$takeover\!\_skill$    & 0.0906  & 1.2399e-54 & &$assist\!\_f\!re$    & 0.0615 &  2.1729e-25 \\
$age$   & 0.0799 &  9.9901e-37 & &$RT\!A$  &  0.0572 &  1.3316e-19 \\
$trust$   & 0.0746 &  2.6316e-25 & &$st\!yle\!\_angry$  & 0.0475 &  2.3371e-03 \\

\bottomrule
\end{tabular}
\end{table}

\subsubsection{Perceived risk}

Table~\ref{tab: sq risk} summarizes the performance of XGBoost-based models in explaining variations in $subjQ_{risk}$ using different input combinations. The baseline model, XGBSQrisk$_{dc}$, shows relatively high RMSE and MAE. Incorporating SA (XGBSQrisk$_{dc+sa}$) leads to substantial improvements, reducing RMSE by 13.03\% and MAE by 12.69\% compared to the baseline ($p_{adjusted} < 0.01$). Adding SC (XGBSQrisk$_{dc+sc}$) results in greater improvements, with RMSE and MAE decreasing by 25.94\% and 25.46\% respectively ($p_{adjusted} < 0.01$). The combined model, XGBSQrisk$_{dc+sa+sc}$, demonstrates the best overall performance among the four models, achieving a 26.26\% reduction in RMSE and a 26.04\% reduction in MAE compared to the baseline ($p_{adjusted} < 0.01$). However, its improvement over XGBSQrisk$_{dc+sc}$ is not statistically significant ($p_{adjusted} = 1.00$). The results indicate that the additional explanatory value provided by SA is minimal when SC is already included, suggesting that SA's contribution largely derives from its overlap with SC, while SC plays a more significant role in explaining $subjQ_{risk}$.



\begin{table}[htbp]
\centering
\caption{Performance of XGBoost-based $subjQ_{risk}$ models.}
\label{tab: sq risk}
\begin{tabular}{llcc}
\toprule
\textbf{Model}& \textbf{Inputs} & \textbf{RMSE ($\downarrow$)} & \textbf{MAE ($\downarrow$)}  \\ \midrule
XGBSQrisk$_{dc}$& $D\!C$  & 1.1348 &	0.8931  \\ 

XGBSQrisk$_{dc+sa}$ & $D\!C$ $+$ $S\!A$  & 0.9869 &	0.7798  \\ 

XGBSQrisk$_{dc+sc}$ & $D\!C$ $+$ $S\!C$ & 0.8404	& 0.6657 \\ 

XGBSQrisk$_{dc+sa+sc}$ & $D\!C$ $+$ $S\!A$ $+$ $S\!C$  & 0.8368 &	0.6605\\ 
\bottomrule
\end{tabular}
\end{table}

\begin{table}[htbp]
\centering
\caption{Feature importance and Bonferroni test results ($p_{adjusted}$) of significant factors of $subjQ_{risk}$.}
\label{tab: risk significant factors}
\begin{tabular}{lcc l l c c}
\toprule
\textbf{Feature} & \textbf{Importance ($\uparrow$)} & \textbf{$p_{adjusted}$ ($\downarrow$)} & &\textbf{Feature} & \textbf{Importance ($\uparrow$)} & \textbf{$p_{adjusted}$ ($\downarrow$)} \\ \midrule

$S\!C$ & 0.3021 & 4.9215e-63 && $accu\!\_dis$ & 0.0360 & 2.9284e-13 \\
$SC_{TC}$ & 0.1812 & 3.5442e-39 && $st\!yle\!\_angry$ & 0.0293 & 2.8887e-09 \\
$trust$ & 0.0672 & 1.0007e-29 && $age$ & 0.0281 & 2.9394e-12 \\
$takeover\!\_skill$ & 0.0607 & 4.7243e-22 && $assist\!\_f\!re$ & 0.0280 & 1.8195e-10 \\
$driving\!\_skill$ & 0.0528 & 4.3703e-38 && $gender$ & 0.0253 & 7.1162e-07 \\
$st\!yle\!\_reckless$ & 0.0482 & 2.2284e-19 && $RT\!A$ & 0.0240 & 4.5956e-05 \\
$st\!yle\!\_anxious$ & 0.0479 & 3.6934e-16 && $accu\!\_years$ & 0.0237 & 9.7020e-07 \\
$SC_{TD}$ & 0.0456 & 2.3842e-16 && && \\

\bottomrule
\end{tabular}
\end{table}


This study delves into the importance of specific DC and SC factors in determining $subjQ_{risk}$. Bonferroni tests identify 16 factors (12 DC-related and 4 SC-related) with strong statistical significance ($p_{adjusted} < 0.05$), which are fed into an XGBSQrisk model. Table \ref{tab: risk significant factors} presents the feature importance and Bonferroni-adjusted significance. Among these factors, subjective SC-related factors, particularly drivers' perceived takeover abilities (as reflected in $S\!C$ and $SC_{TC}$), emerge as dominant contributors. This underscores the central role of self-assessed competence and confidence in shaping drivers' risk perception. A summary plot derived from SHAP values is illustrated in Figure \ref{fig: sufficiency summary plot}. The results align with the feature importance findings, confirming that $S\!C$ and $SC_{TC}$ are dominant determinants of $subjQ_{risk}$. Specifically, the higher the drivers' confidence in fulfilling the takeover tasks (i.e., higher values of $S\!C$ and $SC_{TC}$), the lower their perceived risk. Besides, these two factors are followed by $driving\!\_skill$ and $takeover\_skil$, which are also competence-related factors. The SHAP values of $driving\!\_skill$ and $takeover\_skil$ indicate that drivers with higher levels of skill tend to perceive lower levels of risk across takeover scenarios, reinforcing the importance of competence in shaping subjective risk perception. 


\begin{figure*}[!h]
    \centering
    \includegraphics[width=0.7\columnwidth]{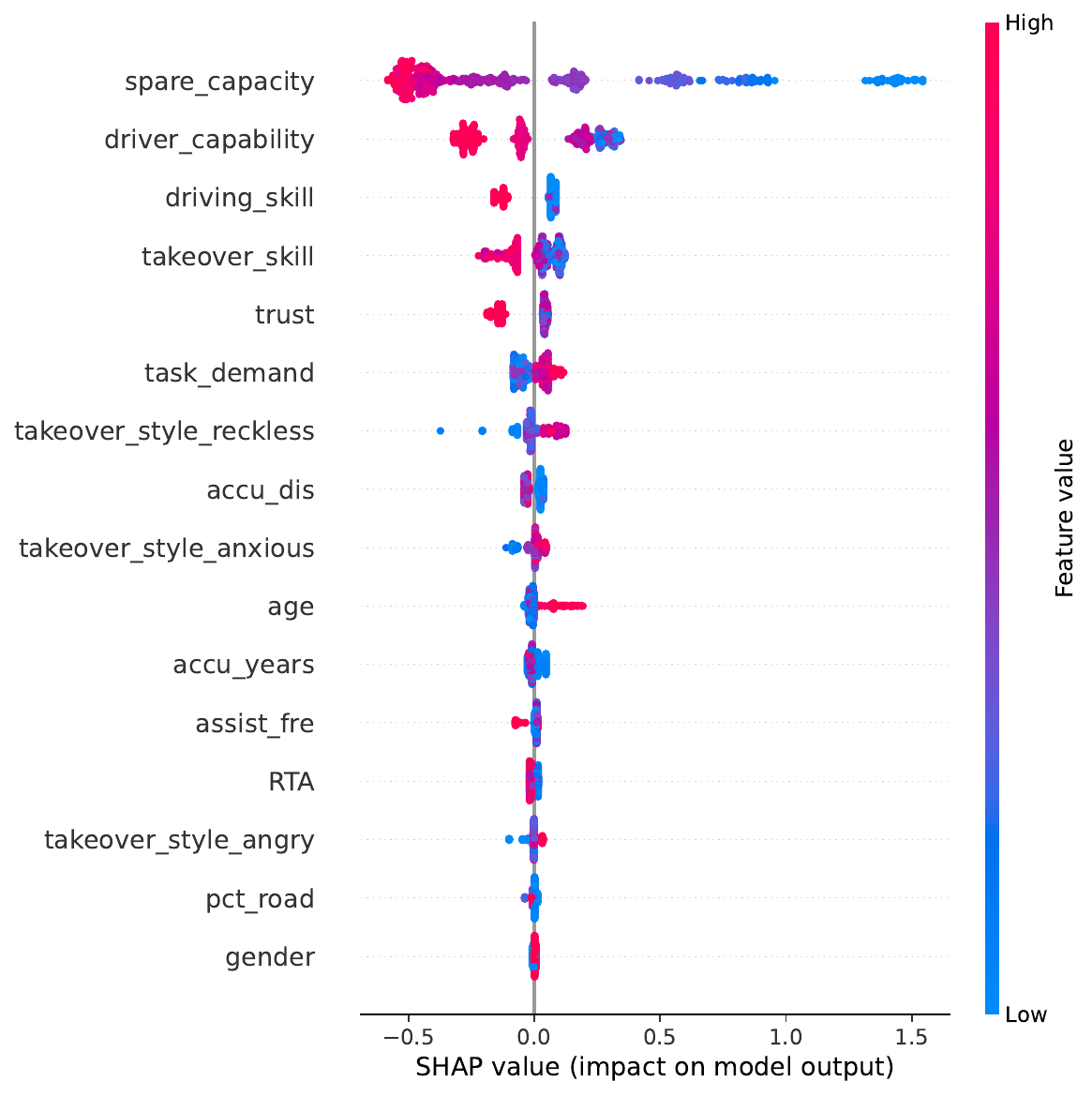}
    \caption{Summary plot of significant Driver Characteristics ($D\!C$) and Spare Capacity ($S\!C$) factors influencing drivers' perceived risk ($subjQ_{risk}$).}
    \label{fig: risk summary plot}
\end{figure*}

\subsubsection{Performance satisfaction}

Table~\ref{tab: sq satis} summarizes the performance of XGBoost-based models in explaining variations in $subjQ_{satis\!f\!action}$ using different input combinations. The baseline model, XGBSQsatis$_{dc}$, exhibits relatively high RMSE and MAE. Adding SA (XGBSQsatis$_{dc+sa}$) significantly reduces RMSE by 16.26\% and MAE by 16.65\% ($p_{adjusted} < 0.01$). Incorporating SC (XGBSQsatis$_{dc+sc}$) leads to more substantial improvements, with RMSE and MAE decreasing by 37.10\% and 36.21\% respectively ($p_{adjusted} < 0.01$), compared to the baseline. Notably, the combined model, XGBSQsatis$_{dc+sa+sc}$, shows no significant improvement over XGBSQsatis$_{dc+sc}$ ($p_{adjusted} = 1.00$), indicating that the additional value provided by SA becomes negligible when SC is included. This suggests that  SC encompasses more unique and valuable information for explaining $subjQ_{satis\!f\!action}$, extending beyond the commonalities it shares with SA. 

\begin{table}[htbp]
\centering
\caption{Performance of XGBoost-based $subjQ_{satis\!f\!action}$ models.}
\label{tab: sq satis}
\begin{tabular}{llcc}
\toprule
\textbf{Model}& \textbf{Inputs} & \textbf{RMSE ($\downarrow$)} & \textbf{MAE ($\downarrow$)}  \\ \midrule
XGBSQsatis$_{dc}$& $D\!C$  & 1.0436	& 0.7771 \\ 

XGBSQsatis$_{dc+sa}$ & $D\!C$ $+$ $S\!A$  & 0.8739	& 0.6477  \\ 

XGBSQsatis$_{dc+sc}$ & $D\!C$ $+$ $S\!C$ & 0.6564	& 0.4957 \\ 

XGBSQsatis$_{dc+sa+sc}$ & $D\!C$ $+$ $S\!A$ $+$ $S\!C$  & 0.6593	& 0.4951\\ 
\bottomrule
\end{tabular}
\end{table}

\begin{figure*}[!h]
    \centering
    \includegraphics[width=0.7\columnwidth]{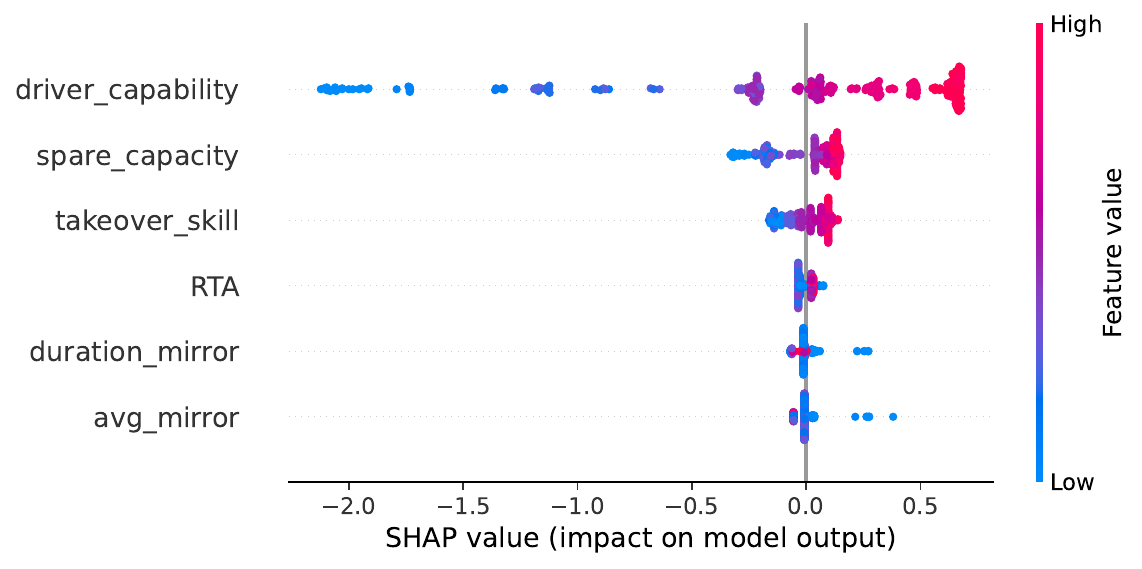}
    \caption{Summary plot of significant Driver Characteristics ($D\!C$) and Spare Capacity ($S\!C$)- factors influencing drivers' perceived performance satisfaction ($subjQ_{satis\!f\!action}$).}
    \label{fig: satisfaction summary plot}
\end{figure*}

This study examines the role of specific DC- and SC-related factors in influencing $subjQ_{satis\!f\!action}$. Using Bonferroni tests, six statistically significant determinants ($p_{adjusted} < 0.05$) are identified and integrated into an XGBSQsatis model. Table \ref{tab: satisfaction significant factors} displays the feature importance and Bonferroni-adjusted significance. The results demonstrate that SC-related factors contribute more significantly to $subjQ_{satis\!f\!action}$ compared to DC-related factors, with subjective SC factors playing a more prominent role than objective SC factors. Similar to the findings on drivers' risk perception, SC-related factors—particularly drivers' perceived confidence in takeover tasks ($SC_{TC}$ and $S\!C$)—emerge as the most dominant determinants of $subjQ_{satis\!f\!action}$. This finding underscores a correlation between drivers' perceived risk and their satisfaction with performance, as drivers are likely to feel dissatisfied with their performance during takeovers if they perceive higher levels of risk. The summary plot in Figure \ref{fig: satisfaction summary plot} reveals that drivers who exhibit superficial mirror-checking behaviour (both in terms of average fixation and total duration) combined with a confident, risk-embracing attitude and high self-perceived takeover skills are more likely to report higher satisfaction with their takeover performance.

\begin{table}[htbp]
\centering
\caption{Feature importance and Bonferroni test results ($p_{adjusted}$) of significant factors of $subjQ_{satis\!f\!action}$.}
\label{tab: satisfaction significant factors}
\begin{tabular}{lcc l l c c}
\toprule
\textbf{Feature} & \textbf{Importance ($\uparrow$)} & \textbf{$p_{adjusted}$ ($\downarrow$)} & &\textbf{Feature} & \textbf{Importance ($\uparrow$)} & \textbf{$p_{adjusted}$ ($\downarrow$)} \\ \midrule

$SC_{TC}$&0.5512 &7.5965e-74 &&$AVG_{mirror}$&0.0792 &8.4662e-11 \\ 
$S\!C$&0.1933 &4.6711e-21 &&$takeover\!\_skill$&0.0592 &8.6937e-26 \\ 
$DU\!R_{mirror}$&0.0879 &2.8561e-08 &&$RT\!A$&0.0292 &1.9477e-02 \\ 

\bottomrule
\end{tabular}
\end{table}


\vspace{-2em}
\subsubsection{Summary}

Three main patterns in drivers' subjective takeover experience are identified: \begin{inparaenum}[(i)]
    \item Perceived time sufficiency is primarily explained by stable driver traits;
    \item Spare Capacity (SC) is the dominant factor for both risk perception and satisfaction; while
    \item Situational Awareness (SA) provides limited additional value when SC is already considered.
\end{inparaenum}




\subsection{Objective Takeover Quality}
\label{subsec: objective quality}

In this section, we examine the factors that influence objective takeover quality indicated by operational data. To capture objective takeover quality, four key metrics spanning the period from the initiation of the takeover request to the lane change are selected: \begin{inparaenum}[(i)]
    \item the minimum time to collision ($objQ_{ttc}$),
    \item the maximum steering wheel angle ($objQ_{steer}$),
    \item the maximum acceleration ($objQ_{acc}$), and
    \item the maximum deceleration ($objQ_{dec}$).
\end{inparaenum}


\subsubsection{Minimum time to collision}

Table~\ref{tab: oq ttc} summarizes the performance of XGBoost-based models in explaining variations in $objQ_{ttc}$ using different predictors. The results indicate no statistically significant improvements in model performance with the inclusion of SA or SC ($p_{adjusted} = 1.00$) when compared to the baseline model XGBOQttc$_{dc}$. Meanwhile, the combined model XGBOQttc$_{dc+sa+sc}$ shows marginal improvement over the baseline model, though this difference is not statistically significant ($p_{adjusted} = 1.00$). These results suggest that the addition of SA or SC does not offer substantial explanatory power beyond the basic driver characteristics (DC).


\begin{table}[htbp]
\centering
\caption{Performance of XGBoost-based $objQ_{ttc}$ models.}
\label{tab: oq ttc}
\begin{tabular}{llcc}
\toprule
\textbf{Model}& \textbf{Inputs} & \textbf{RMSE ($\downarrow$)} & \textbf{MAE ($\downarrow$)}  \\ \midrule
XGBOQttc$_{dc}$& $D\!C$  & 1.0407	& 0.8175  \\ 

XGBOQttc$_{dc+sa}$ & $D\!C$ $+$ $S\!A$  & 1.0486	& 0.8212  \\ 

XGBOQttc$_{dc+sc}$ & $D\!C$ $+$ $S\!C$ & 1.0099	& 0.7979 \\ 

XGBOQttc$_{dc+sa+sc}$ & $D\!C$ $+$ $S\!A$ $+$ $S\!C$  & 1.0093	& 0.7980\\ 
\bottomrule
\end{tabular}
\end{table}

\begin{figure*}[!h]
    \centering
    \includegraphics[width=0.7\columnwidth]{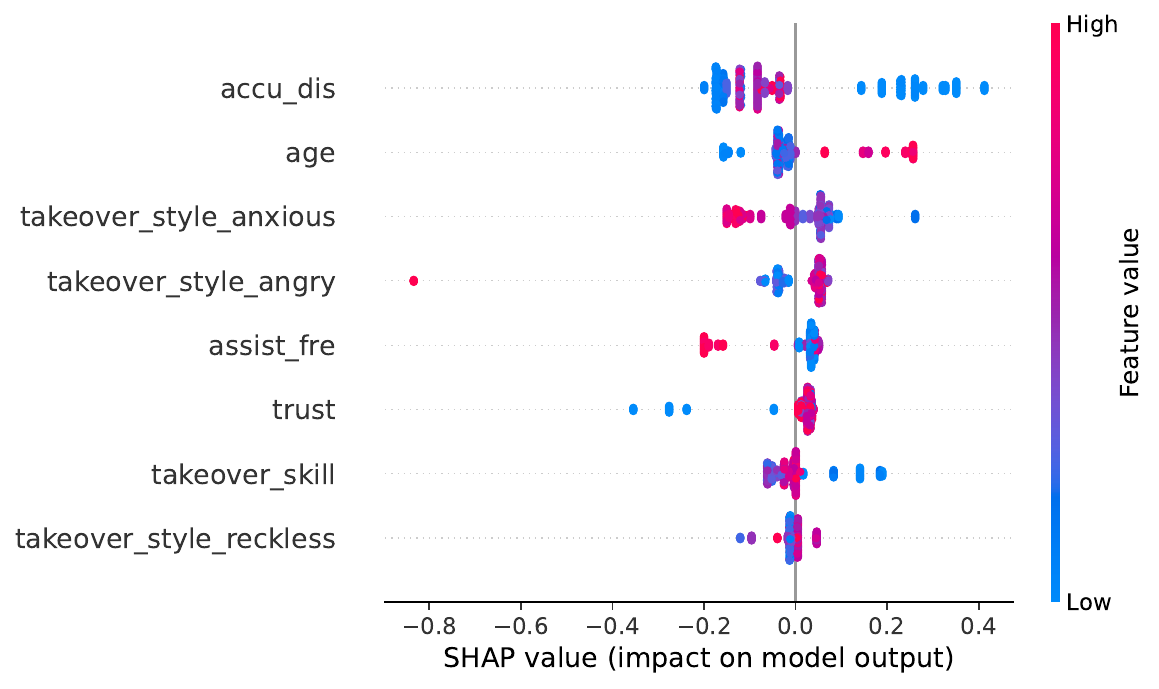}
    \caption{Summary plot of significant Driver Characteristics ($D\!C$) factors influencing minimum time to collision ($objQ_{ttc}$).}
    \label{fig: ttc summary plot}
\end{figure*}

This study investigates the influence of key DC-related factors on $objQ_{ttc}$ in response to a takeover request. Through Bonferroni-adjusted significance tests, eight significant factors are identified ($p_{adjusted} < 0.01$) and incorporated into an XGBoost-based $objQ_{ttc}$ model to assess their relative contributions. As shown in Table \ref{tab: ttc significant factors}, $accu\!\_dis$ and $age$ are the two most influential factors, suggesting that overall driving experience plays a crucial role in affecting $objQ_{ttc}$. These two objective proxies for driving experience outweigh self-reported $takeover\!\_skill$ in modeling $objQ_{ttc}$, suggesting that for objective performance metrics like $objQ_{ttc}$, objective measures may carry greater predictive power than subjective self-assessments. Besides, the summary plot in Figure \ref{fig: ttc summary plot} illustrates that, drivers who exhibit longer minimum time to collision tend to be older, use driver assistance systems less frequently, have a less anxious takeover style, and demonstrate high trust in conditional driving automation. Besides, inexperienced drivers (low in $accu\!\_dis$ and $takeover\!\_skill$) fall into two distinct groups. Some compensate for their lack of experience by adopting a cautious and conservative takeover strategy, maintaining a shorter $objQ_{ttc}$. Others, however, may struggle with reaction timing, leading to more abrupt takeovers that results in longer $objQ_{ttc}$.



\begin{table}[htbp]
\centering
\caption{Feature importance and Bonferroni test results ($p_{adjusted}$) of significant factors of $objQ_{ttc}$.}
\label{tab: ttc significant factors}
\begin{tabular}{lcc l l c c}
\toprule
\textbf{Feature} & \textbf{Importance ($\uparrow$)} & \textbf{$p_{adjusted}$ ($\downarrow$)} & &\textbf{Feature} & \textbf{Importance ($\uparrow$)} & \textbf{$p_{adjusted}$ ($\downarrow$)} \\ \midrule

$accu\!\_dis$ & 0.1727 & 4.2915e-41 & & $takeover\!\_skill$  & 0.1173 & 5.2461e-12 \\
$age$ & 0.1544 & 1.3432e-27 & & $assist\!\_f\!re$  & 0.1124 & 1.6448e-17 \\
$st\!yle\!\_reckless$ & 0.1387 & 2.1205e-12 & & $st\!yle\!\_anxious$ & 0.1030 & 5.0132e-14 \\
$st\!yle\!\_angry$  & 0.1174 & 2.3469e-27 & & $trust$ & 0.0841 & 1.3646e-21 \\

\bottomrule
\end{tabular}
\end{table}




\vspace{-2em}
\subsubsection{Maximum steering wheel angle}

Table~\ref{tab: oq steer} presents the performance of XGBoost-based models in explaining variations in $objQ_{steer}$ using different predictors. The baseline model, XGBOQsteer$_{dc}$, establishes a solid foundation with low RMSE and MAE. Adding SA alone (XGBOQsteer$_{dc+sa}$) results in minor performance gains, reducing RMSE by 3.44\% and MAE by 3.90\%, but these improvements are not statistically significant ($p_{adjusted} > 0.05$). In contrast, incorporating SC (XGBOQsteer$_{dc+sc}$) significantly enhances model performance, reducing RMSE by 10.03\% and MAE by 18.25\% ($p_{adjusted} < 0.05$). The combined model, XGBOQsteer$_{dc+sa+sc}$, yields performance metrics nearly identical to XGBOQsteer$_{dc+sc}$, indicating that SA adds minimal value when DC and SC are accounted for. 





\begin{table}[htbp]
\centering
\caption{Performance of XGBoost-based $objQ_{steer}$ models.}
\label{tab: oq steer}
\begin{tabular}{llcc}
\toprule
\textbf{Model}& \textbf{Inputs} & \textbf{RMSE ($\downarrow$)} & \textbf{MAE ($\downarrow$)}  \\ \midrule
XGBOQsteer$_{dc}$& $D\!C$  & 0.3580	& 0.1874  \\ 

XGBOQsteer$_{dc+sa}$ & $D\!C$ $+$ $S\!A$  & 0.3457 &	0.1801  \\ 

XGBOQsteer$_{dc+sc}$ & $D\!C$ $+$ $S\!C$ & 0.3221	& 0.1532 \\ 

XGBOQsteer$_{dc+sa+sc}$ & $OQDC$ $+$ $S\!A$ $+$ $S\!C$  & 0.3253	& 0.1542\\ 
\bottomrule
\end{tabular}
\end{table}

\begin{figure*}[!h]
    \centering
    \includegraphics[width=0.7\columnwidth]{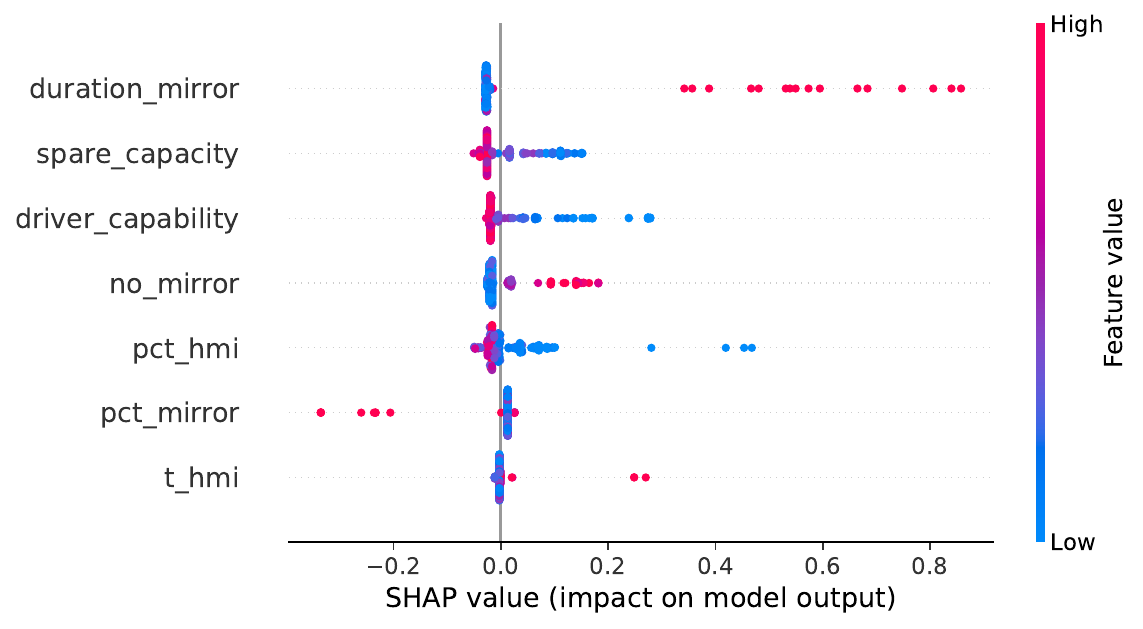}
    \caption{Summary plot of significant Spare Capacity ($S\!C$) factors influencing maximum steering wheel angle ($objQ_{steer}$).}
    \label{fig: steer summary plot}
\end{figure*}

We explore the contributions of DC- and SC-related factors in determining $objQ_{steer}$. Bonferroni tests identify seven SC-related factors as statistically significant predictors ($p_{adjusted} < 0.01$), with their relative importance and significances detailed in Table \ref{tab: steer significant factors}. Two objective gaze metrics—$DU\!R_{mirror}$ and $NO_{mirror}$—exhibit greater importance than self-reported SC factors in explaining steering behaviours. Figure \ref{fig: steer summary plot} further reveals that drivers with lower SC - evidenced by increased side mirror checks, delayed first HMI glances, and lower self-reported SC ratings - generally exhibit larger maximum steering wheel angles during lane-changes. Besides, we notice that certain factors have long tails in their SHAP distributions, particularly for $DU\!R_{mirror}$, reflecting their spread impacts on $objQ_{steer}$. 






\begin{table}[htbp]
\centering
\caption{Feature importance and Bonferroni test results ($p_{adjusted}$) of significant factors of $objQ_{steer}$.}
\label{tab: steer significant factors}
\begin{tabular}{lcc l l c c}
\toprule
\textbf{Feature} & \textbf{Importance ($\uparrow$)} & \textbf{$p_{adjusted}$ ($\downarrow$)} & &\textbf{Feature} & \textbf{Importance ($\uparrow$)} & \textbf{$p_{adjusted}$ ($\downarrow$)} \\ \midrule

$DU\!R_{mirror}$ & 0.3438  &  2.1701e-08 &  & $SC_{TC}$ & 0.1116 & 5.4067e-07 \\
$NO_{mirror}$ & 0.2409  &  1.6282e-06 &  & $t_{HMI}$ & 0.0864 & 1.3851e-04 \\
$S\!C$ & 0.1315 & 4.7475e-18 &  & $st\!yle\!\_anxious$ & 0.0859 & 8.8415e-07 \\

\bottomrule
\end{tabular}
\end{table}


\vspace{-2em}
\subsubsection{Maximum acceleration}

Table~\ref{tab: oq acc} presents the performance comparison of XGBoost-based models in explaining variations in $objQ_{acc}$ using different predictors. The baseline model, XGBOQacc$_{dc}$, achieves the highest RMSE and MAE. Incorporating SA (XGBOQacc$_{dc+sa}$) slightly reduces both RMSE (by 0.73\%) and MAE (by 1.79\%), but these improvements are not statistically significant ($p_{adjusted} > 0.05$). In contrast, adding SC (XGBOQacc$_{dc+sc}$) results in substantial performance enhancements, with RMSE reduced by 13.94\% and MAE by 17.68\%, both statistically significant ($p_{adjusted} < 0.01$). The combined model (XGBOQacc$_{dc+sa+sc}$) performs comparably to XGBOQacc$_{dc+sc}$, showing only marginal, non-significant differences ($p_{adjusted} > 0.05$). These findings emphasize SC's pivotal role in enhancing model performance, while the incremental value of SA is minimal when DC and SC are already included.


\begin{table}[htbp]
\centering
\caption{Performance of XGBoost-based $objQ_{acc}$ models.}
\label{tab: oq acc}
\begin{tabular}{llcc}
\toprule
\textbf{Model}& \textbf{Inputs} & \textbf{RMSE ($\downarrow$)} & \textbf{MAE ($\downarrow$)}  \\ \midrule
XGBOQacc$_{dc}$& $D\!C$  & 1.6226	& 1.2291  \\ 

XGBOQacc$_{dc+sa}$ & $D\!C$ $+$ $S\!A$  & 1.6107	& 1.2071  \\ 

XGBOQacc$_{dc+sc}$ & $D\!C$ $+$ $S\!C$ & 1.3964	& 1.0118 \\ 

XGBOQacc$_{dc+sa+sc}$ & $D\!C$ $+$ $S\!A$ $+$ $S\!C$  & 1.4076	& 1.0224\\ 
\bottomrule
\end{tabular}
\end{table}

\begin{figure*}[!h]
    \centering
    \includegraphics[width=0.7\columnwidth]{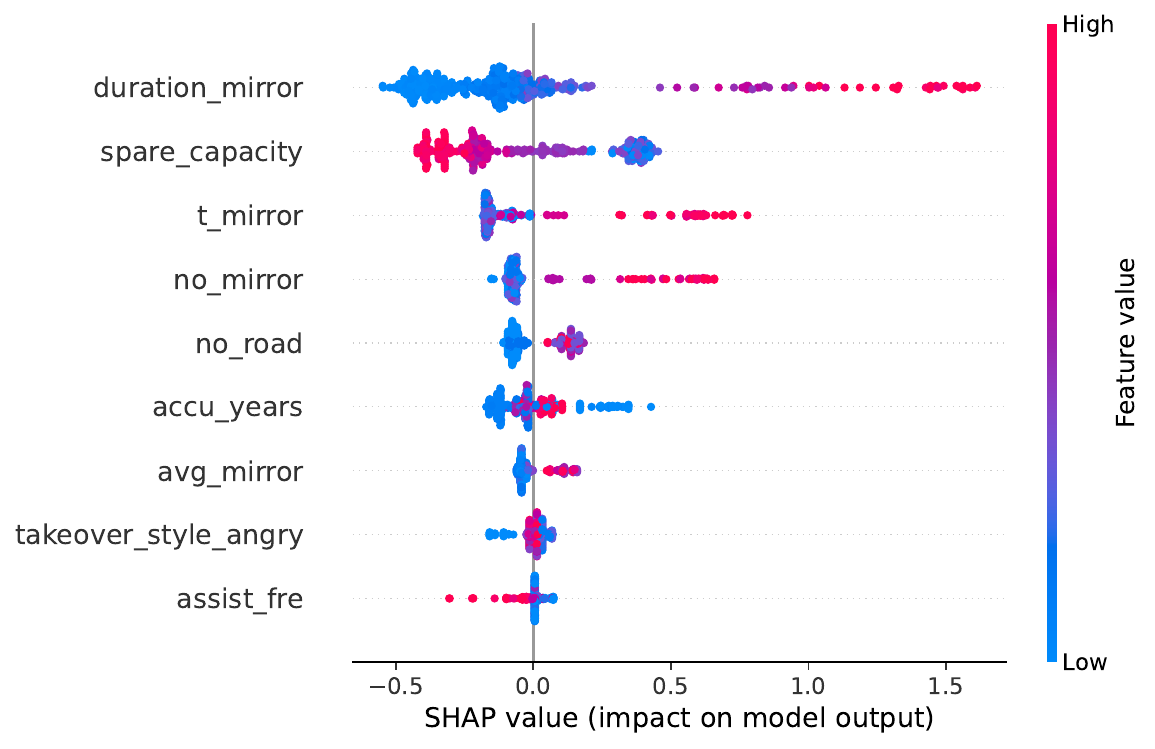}
    \caption{Summary plot of significant Driver Characteristics ($D\!C$) and Spare Capacity ($S\!C$) factors influencing maximum acceleration ($objQ_{acc}$).}
    \label{fig: acc summary plot}
\end{figure*}

This study examines the importance of DC- and SC-related factors in modeling drivers' maximum acceleration ($objQ_{acc}$) in response to a takeover request. Bonferroni tests identify three DC-related factors and six SC-related factors (including five objective measurements and one subjective measurement) as significant determinants of $objQ_{acc}$ ($p_{adjusted} < 0.05$). Their feature importance and statistical significances are summarized in Table \ref{tab: acc significant factors}. Overall, SC-related factors are the dominant determinants of $objQ_{acc}$, while DC-related factors contribute significantly but to a lesser extent. Among SC-related factors, objective gaze metrics are more influential than self-reported SC ratings. Four most important features — $DU\!R_{mirror}$, $NO_{mirror}$, $AVG_{mirror}$, and $t_{mirror}$ — suggest that side mirrors are the most critical AOI for modeling $objQ_{acc}$. 
Further, as shown in the summary plot in Figure \ref{fig: acc summary plot}, drivers with lower SC in fulfilling the takeover task - as indicated by allocating greater visual attention on side mirrors and the forward road, taking longer time to first check side mirrors, and reporting perception of lower SC - tend to exhibit larger maximum accelerations in response to takeover requests. 



\begin{table}[htbp]
\centering
\caption{Feature importance and Bonferroni test results ($p_{adjusted}$) of significant factors of $objQ_{acc}$.}
\label{tab: acc significant factors}
\begin{tabular}{lccllcc}
\toprule
\textbf{Feature} & \textbf{Importance ($\uparrow$)} & \textbf{$p_{adjusted}$ ($\downarrow$)} & &\textbf{Feature} & \textbf{Importance ($\uparrow$)} & \textbf{$p_{adjusted}$ ($\downarrow$)} \\ \midrule

$DU\!R_{mirror}$ & 0.2363 & 1.4877e-32 &  & $accu\!\_years$ & 0.0608 & 1.4354e-25 \\
$NO_{mirror}$ & 0.2251 & 9.6136e-14 &  & $NO_{road}$ & 0.0486 & 1.9576e-05 \\
$AVG_{mirror}$ & 0.1247 & 1.6751e-02 &  & $assist\!\_f\!re$ & 0.0436 & 8.6914e-05 \\
$t_{mirror}$ & 0.1206 & 5.4337e-20 &  & $st\!yle\!\_angry$ & 0.0377 & 4.8701e-14 \\
$S\!C$ & 0.1027 & 2.6390e-37 &  &  &  &  \\

\bottomrule
\end{tabular}
\end{table}


\subsubsection{Maximum deceleration}

Table~\ref{tab: oq dec} presents the performance of XGBoost-based models in explaining $objQ_{dec}$ with various predictors. The baseline model, XGBOQdec$_{dc}$, achieves the highest RMSE and MAE. Adding SA (XGBOQdec$_{dc+sa}$) reduces RMSE by 2.61\% and MAE by 2.85\%, but these improvements are not statistically significant ($p_{adjusted} > 0.05$). In contrast, incorporating SC (XGBOQdec$_{dc+sc}$) substantially enhances model performance by reducing RMSE by 13.14\% and MAE by 15.78\% ($p_{adjusted} < 0.01$) when compared to the baseline model. The combined model (XGBOQdec${dc+sa+sc}$) performs almost identically to XGBOQdec${dc+sc}$, with no statistically significant differences ($p_{adjusted} > 0.05$). These findings emphasize the crucial contribution of SC in explaining variations in $objQ_{dec}$, while the addition of SA offers only minimal incremental value. The RMSE and MAE values of the $objQ_{dec}$ models are relatively high compared to those of the other takeover performance metrics. This suggests that drivers' deceleration responses to a takeover request are less predictable, and there may be additional factors influencing maximum deceleration during takeovers that are not fully captured by the current predictors.


\begin{table}[htbp]
\centering
\caption{Performance of XGBoost-based $objQ_{dec}$ models.}
\label{tab: oq dec}
\begin{tabular}{llcc}
\toprule
\textbf{Model}& \textbf{Inputs} & \textbf{RMSE ($\downarrow$)} & \textbf{MAE ($\downarrow$)}  \\ \midrule
XGBOQdec$_{dc}$& $D\!C$  & 3.3894 &	2.9508  \\ 

XGBOQdec$_{dc+sa}$ & $D\!C$ $+$ $S\!A$  & 3.3009	& 2.8667 \\ 

XGBOQdec$_{dc+sc}$ & $D\!C$ $+$ $S\!C$ & 2.9440	& 2.4852 \\ 

XGBOQdec$_{dc+sa+sc}$ & $D\!C$ $+$ $S\!A$ $+$ $S\!C$  & 2.9490	& 2.4976\\ 
\bottomrule
\end{tabular}
\end{table}

\begin{figure*}[!h]
    \centering
    \includegraphics[width=0.7\columnwidth]{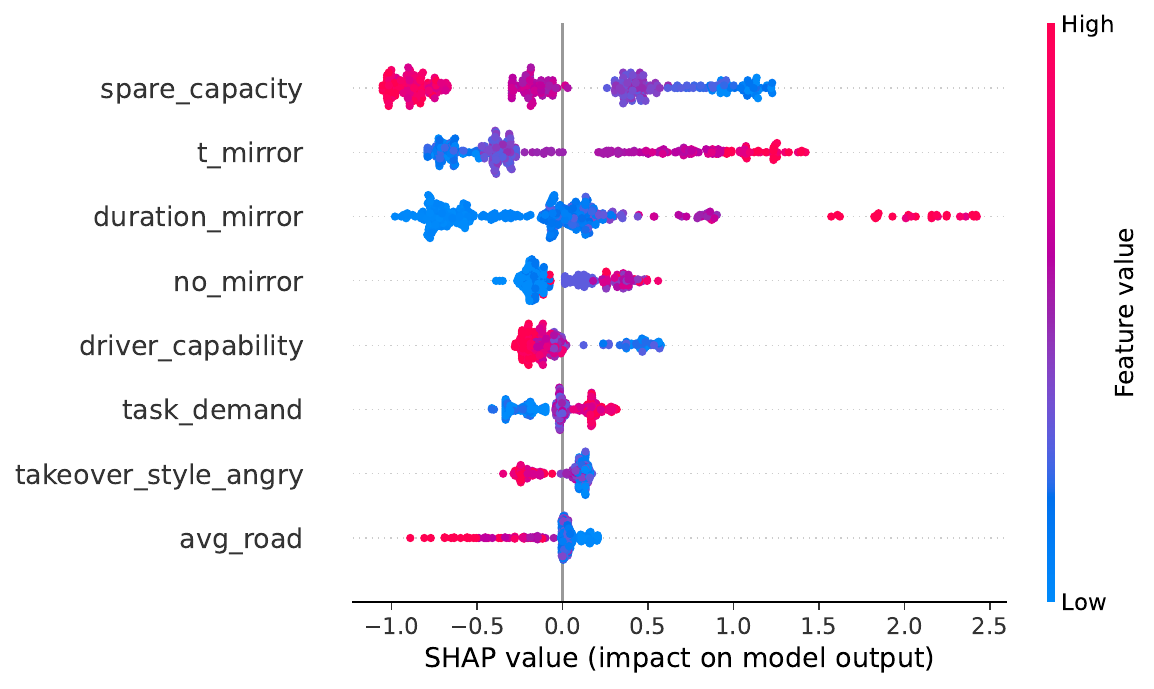}
    \caption{Summary plot of significant Spare Capacity ($S\!C$) factors influencing maximum deceleration ($objQ_{dec}$).}
    \label{fig: dec summary plot}
\end{figure*}

We investigate the feature importance and Bonferroni-adjusted significance results for individual DC- and SC-related factors influencing $objQ_{dec}$. The results, as summarized in Table \ref{tab: dec significant factors}, identify eight significant factors ($p_{adjusted}$ < 0.01), with seven SC-related factors playing a dominant role in modeling $objQ_{dec}$. Specifically, self-reported $S\!C$ and $SC_{TC}$ are the two most influential determinants, followed by two gaze metrics related to side mirrors ($DU\!R_{mirror}$ and $t_{mirror}$). Besides, Figure \ref{fig: dec summary plot} reveals that drivers who exhibit larger maximum deceleration in response to takeover requests generally perceive themselves as having lower SC (i.e., lower driver capability and/or higher task demand). This reduced SC can also be reflected in their gaze behaviours, as they take longer to first check side mirrors and allocate more visual attention to them (more times and longer total duration). 



\begin{table}[htbp]
\centering
\caption{Feature importance and Bonferroni test results ($p_{adjusted}$) of significant factors of $objQ_{dec}$.}
\label{tab: dec significant factors}
\begin{tabular}{lcc l l cc}
\toprule
\textbf{Feature} & \textbf{Importance ($\uparrow$)} & \textbf{$p_{adjusted}$ ($\downarrow$)} & &\textbf{Feature} & \textbf{Importance ($\uparrow$)} & \textbf{$p_{adjusted}$ ($\downarrow$)} \\ \midrule

$S\!C$ & 0.2977 & 2.0528e-35 &  & $SC_{TD}$ & 0.0862 & 9.5316e-10 \\
$SC_{TC}$ & 0.2015 & 7.7216e-16 &  & $AVG_{road}$ & 0.0744 & 1.1295e-08 \\
$DU\!R_{mirror}$ & 0.1052 & 2.3073e-34 &  & $NO_{mirror}$ & 0.0697 & 1.3810e-15 \\
$t_{mirror}$ & 0.0992 & 1.6802e-28 &  & $st\!yle\!\_angry$ & 0.0662 & 1.0259e-12 \\

\bottomrule
\end{tabular}
\end{table}

\subsubsection{Summary}

Three main patterns are identified in objective takeover performance request initiation to lane change: \begin{inparaenum}[(i)]
    \item Minimum time-to-collision is predominantly determined by basic Driver Characteristics (DC);
    \item Spare Capacity (SC) significantly influences the remaining three objective measures (maximum steering wheel angle, acceleration, and deceleration), exhibiting an inverse relationship where lower SC corresponds to larger values; while
    \item Situational Awareness (SA) contributes minimally when DC and SC are already accounted for.
\end{inparaenum}


\section{Discussion}
\label{sec: discussions}


To ensure the robustness of our findings across methodologies, we validate XGBoost’s results using two complementary models: Random Forest (RF) and Light Gradient Boosting Machine (LightGBM). RF, as a bagging method, is renowned for its robustness and interpretability, effectively mitigating overfitting through random feature selection and averaging across multiple trees. Meanwhile, LightGBM, a highly efficient gradient-boosting framework, offers a direct comparison to XGBoost due to their shared boosting principles but differing implementations. 

Table \ref{tab: model comparison} compares three models’ performance across ten metrics using different combinations of predictors: Driver Characteristics (DC), Situational Awareness (SA), and Spare Capacity (SC). We observe consistent predictive patterns for all performance metrics when incorporating specific predictors. Specifically,

\begin{itemize}
    \item[1] \textbf{among three reaction times:} drivers' instinctive and immediate responses, such as pressing a mode switch button ($t_{button}$) and reorienting visual attention to the road ($t_{road}$), are largely driven by SA. These reflexive actions occur in a short time frame, relying heavily on the driver’s perception of the environment and their readiness to act. In contrast, reflective responses, such as going back to the conscious driving loop ($T\!oT$), involve more deliberate cognitive processes. These responses require drivers to assess the situation, allocate cognitive resources, and execute appropriate actions, which may demand greater integration of SC. This distinction underscores the interplay between rapid, awareness-driven reactions and slower, cognitively mediated processes, offering a deeper understanding of the cognitive mechanisms at play in takeover scenarios. 

    \item[2] \textbf{among three subjective takeover quality metrics:} the baseline model relying solely on DC predominantly explains perceived time sufficiency, indicating that stable traits of drivers play a central role, with minimal additional contributions from SA and SC. Besides, SC emerges as the most influential factor in explaining subjective perceptions of both risk and satisfaction, demonstrating its critical role in shaping these evaluations. While SA-related factors provide supplementary contributions, their additional value for capturing drivers' perceptions of risk and satisfaction is limited when SC is already included in the model. This suggests that the information captured by SC encompasses much of the variance that SA might explain, highlighting the dominant role of SC in understanding subjective evaluations of driving performance.
    
    \item[3] \textbf{among four objective takeover quality metrics:} for minimum time to collision, neither SA nor SC provides substantial explanatory power beyond the baseline model, likely due to information overlap with DC. However, for the remaining three metrics (maximum steering wheel angle, acceleration, and deceleration), SC demonstrates a significant inverse relationship—lower SC consistently corresponds to more abrupt control operations. Notably, SA contributes minimally when combined with DC and SC, mirroring the patterns observed in subjective takeover quality metrics. This consistency across both objective and subjective measures underscores SC's dominant role in determining takeover execution quality.
    
\end{itemize}



\begin{table}[htbp]
\centering
\caption{Model performance comparison for ten performance metrics.}
\label{tab: model comparison}
\begin{tabular}{llcc cc cc }
\toprule
\multirow{2}{6em}{\textbf{Model output}}& \multirow{2}{6em}{\textbf{Model input}} & \multicolumn{2}{c}{\textbf{Random Forest}} & \multicolumn{2}{c}{\textbf{LightGBM}} & \multicolumn{2}{c}{\textbf{XGBoost}}\\ \cline{3-4} \cline{5-6} \cline{7-8}
 & & \textbf{RMSE ($\downarrow$)} & \textbf{MAE ($\downarrow$)}& \textbf{RMSE ($\downarrow$)} & \textbf{MAE ($\downarrow$)}& \textbf{RMSE ($\downarrow$)} & \textbf{MAE ($\downarrow$)}  \\ \midrule

\multirow{4}{4em}{$t_{button}$}& $D\!C$ & 0.6929 & 0.5018 & 0.6908 & 0.5006 & 0.6848 & 0.4966\\ 

& $D\!C$ $+$ $S\!A$  & 0.6195 & 0.4462 & 0.6292 & 0.4526 & 0.6217 & 0.4474\\ 

& $D\!C$ $+$ $S\!C$ & 0.6808 & 0.4933 & 0.6985 & 0.5017 & 0.6657 & 0.4808\\ 

& $D\!C$ $+$ $S\!A$ $+$ $S\!C$  & 0.6362 & 0.4531 & 0.6197 & 0.4521 & 0.6130 & 0.4440\\ 
\midrule

\multirow{4}{4em}{$t_{road}$}& $D\!C$  &0.7143 & 0.5277 & 0.7120 & 0.5264& 0.6973 & 0.5332\\ 

& $D\!C$ $+$ $S\!A$  &0.6471 & 0.4725 & 0.6701 & 0.4866& 0.6298 & 0.4649\\ 

& $D\!C$ $+$ $S\!C$ &0.7060 & 0.5598 & 0.7349 & 0.5573& 0.6889 & 0.5400\\ 

& $D\!C$ $+$ $S\!A$ $+$ $S\!C$  &0.6469 & 0.4752 & 0.6563 & 0.4812& 0.6273 & 0.4667\\ 
\midrule

\multirow{4}{4em}{$T\!oT$}& $D\!C$  & 1.3555 & 1.0474 & 1.3524 & 1.0445 & 1.3082 & 1.0068\\ 

& $D\!C$ $+$ $S\!A$  & 1.2324 & 0.9588 & 1.3019 & 1.0065 & 1.2348 & 0.9512\\ 

& $D\!C$ $+$ $S\!C$ & 1.1942 & 0.9278 & 1.2135 & 0.9329 & 1.1538 & 0.9006\\ 

& $D\!C$ $+$ $S\!A$ $+$ $S\!C$  & 1.1833 & 0.9179 & 1.1557 & 0.8871 & 1.1249 & 0.8804\\ 
\midrule

\multirow{4}{4em}{$subjQ_{suf\!f\!iciency}$}& $D\!C$ &0.1939 & 0.1425& 0.1936 & 0.1421 & 0.2332 & 0.1810\\ 

& $D\!C$ $+$ $S\!A$  & 0.1875 & 0.1390 & 0.1940 & 0.1469 & 0.2325 & 0.1808\\ 

& $D\!C$ $+$ $S\!C$ & 0.2401 & 0.1818 & 0.1835 & 0.1350 & 0.2240 & 0.1712\\ 

& $D\!C$ $+$ $S\!A$ $+$ $S\!C$  & 0.2431 & 0.1840 & 0.1857 & 0.1387 & 0.2232 & 0.1711\\ 
\midrule
			
\multirow{4}{4em}{$subjQ_{risk}$}& $D\!C$  & 1.1439 & 0.8418 & 1.1425 & 0.8406& 1.1348 & 0.8931\\ 

& $D\!C$ $+$ $S\!A$  &0.9814 & 0.7533 & 0.9937 & 0.7538& 0.9869 & 0.7798\\ 

& $D\!C$ $+$ $S\!C$ &0.8540 & 0.6772 & 0.7773 & 0.5724& 0.8404 & 0.6657\\ 

& $D\!C$ $+$ $S\!A$ $+$ $S\!C$  & 0.8656 & 0.6861 & 0.7796 & 0.5764& 0.8368 & 0.6605\\ 
\midrule

\multirow{4}{4em}{$subjQ_{satis\!f\!action}$}& $D\!C$  &1.0809 & 0.7974 & 1.0788 & 0.7948& 1.0436 & 0.7771\\ 

& $D\!C$ $+$ $S\!A$  & 0.8683 & 0.6356 & 0.9127 & 0.6658 & 0.8739 & 0.6477\\ 

& $D\!C$ $+$ $S\!C$ & 0.6713 & 0.4820 & 0.6919 & 0.5010 & 0.6564 & 0.4957\\ 

& $D\!C$ $+$ $S\!A$ $+$ $S\!C$  & 0.6732 & 0.4831 & 0.6849 & 0.4951 & 0.6593 & 0.4951\\ 
\midrule	

\multirow{4}{4em}{$objQ_{ttc}$}& $D\!C$  & 1.0353 & 0.8062 & 1.0337 & 0.8043 & 1.0407 & 0.8175\\ 

& $D\!C$ $+$ $S\!A$  & 1.0177 & 0.7930 & 1.0739 & 0.8378 & 1.0486 & 0.8212\\ 

& $D\!C$ $+$ $S\!C$ & 1.0623 & 0.8459 & 0.9978 & 0.7694 & 1.0099 & 0.7979\\ 

& $D\!C$ $+$ $S\!A$ $+$ $S\!C$  & 1.0666 & 0.8488 & 1.0006 & 0.7742 & 1.0093 & 0.7980\\ 
\midrule

\multirow{4}{4em}{$objQ_{steer}$}& $D\!C$  & 0.3781 & 0.1965 & 0.3752 & 0.1952 & 0.3580 & 0.1874\\ 

& $D\!C$ $+$ $S\!A$  &0.3769 & 0.1959 & 0.3759 & 0.2087 & 0.3457 & 0.1801\\ 

& $D\!C$ $+$ $S\!C$ &0.3337 & 0.1629 & 0.3382 & 0.1694 & 0.3221 & 0.1532\\ 

& $D\!C$ $+$ $S\!A$ $+$ $S\!C$  & 0.3342 & 0.1631 & 0.3381 & 0.1698 & 0.3253 & 0.1542\\ 
\midrule

\multirow{4}{4em}{$objQ_{acc}$}& $D\!C$  &1.6702 & 1.2345 & 1.6616 & 1.2337 & 1.6226 & 1.2291\\ 

& $D\!C$ $+$ $S\!A$  & 1.6204 & 1.2288 & 1.7325 & 1.2861 & 1.6107 & 1.2071\\ 

& $D\!C$ $+$ $S\!C$ & 1.4761 & 1.0917 & 1.4078 & 1.0132& 1.3964 & 1.0118\\ 

& $D\!C$ $+$ $S\!A$ $+$ $S\!C$  &1.4717 & 1.0887 & 1.4424 & 1.0410 & 1.4076 & 1.0224\\ 
\midrule

\multirow{4}{4em}{$objQ_{dec}$}& $D\!C$  & 3.5099 & 3.0026 & 3.4981 & 2.9978 & 3.3894 & 2.9508\\ 

& $D\!C$ $+$ $S\!A$  & 3.2757 & 2.8108 & 3.4621 & 2.8659 & 3.3009 & 2.8667\\ 

& $D\!C$ $+$ $S\!C$ & 3.0097 & 2.5845 & 3.0476 & 2.4988& 2.9440 & 2.4852\\ 

& $D\!C$ $+$ $S\!A$ $+$ $S\!C$  &  3.0173 & 2.5921 & 2.9949 & 2.4487 & 2.9490 & 2.4976\\

\bottomrule
\end{tabular}
\end{table}


The consistency in predictive performance across all three models reinforces the robustness of our findings that drivers' reflexive responses to takeover requests are largely influenced by their SA, and the quality of takeover performance is largely determined by their SC. This consistency suggests that the observed patterns are not model-specific but reflect underlying relationships in the data.

Besides, our finding aligns closely with prior studies. A strong correlation has been observed between drivers' SA and their reflexive reactions to takeover requests. For instance, \cite{zeeb2015determines} demonstrated that drivers who were more aware of their surroundings (higher SA) took shorter time to put their hands back on the steering wheel. \cite{clark2017situational} found that lower SA led to delays in pressing a takeover button to indicate that drivers wanted to regain vehicle control. However, when it comes to takeover time which requires more cognitive resources than reflexive reactions, the role of SA becomes less straightforward. \cite{jia2024drivers} revealed that higher SA was not always associated with shorter takeover time, especially in complex situations. It is important to note that some studies report a link between improved SA and reduced takeover time, but these investigations often do not differentiate between reflexive reaction time and takeover time, leading to inconsistent conclusions regarding the role of SA. \cite{jia2024drivers} also reported that elevated SA does not necessarily enhance takeover quality, as measured by parameters such as maximum longitudinal and lateral acceleration, minimum time-to-collision, and maximum road deviation. Similarly, \cite{agrawal2021evaluating} found that drivers' SA before takeover does not significantly affect takeover quality metrics (e.g., collision risk, intensity of driver response, and trajectory quality), whereas increased mental stress detrimentally impacts these outcomes. \cite{liu2024safety} identified cognitive load as a critical predictor of takeover safety. Collectively, these studies suggest that cognition-related constructs—such as spare capacity and mental load—play a more decisive role in determining takeover quality than SA alone. This convergence of evidence not only reinforces the validity of our findings but also highlights the importance of monitoring driver cognitive states (such as SA and SC) to adjust interaction strategies and optimize driver performance during vehicle control transitions. These insights carry significant implications for the design of advanced driver-assistance systems, which can leverage these findings to more accurately estimate driver reactions and implement personalized, adaptive alert strategies.



This study is subject to three primary limitations:
\begin{inparaenum}[(i)]
\item The experiment was conducted in a simulated environment rather than in real-world driving conditions. While simulators offer controlled settings that facilitate experimentation, driver behaviour may differ in actual driving situations, potentially limiting the generalizability of our findings; 
\item The analysis primarily examines the impact of individual factors and factor groups on takeover performance. Although SHAP values provide a preliminary insight into the interactions among these factors, we do not fully explore their interplays. A more in-depth investigation, especially into the interplay between situational awareness and spare capacity, could uncover nuanced dynamics within vehicle control transitions; and 
\item This study does not examine the correlation between SA and SC, as the focus is on understanding their distinct roles in influencing different aspects of takeover performance. However, a deeper investigation into the interrelationship between SA and SC and the extent of their unique vs. overlapping contributions could provide further insights into their respective roles in driver behaviour during takeover scenarios. 
\end{inparaenum}
Addressing these limitations could enhance the validity, applicability, and practical relevance of the findings to real-world driving contexts. Such efforts would also help identify more robust strategies for improving the safety and effectiveness of human interactions with automated driving systems.

\section{Conclusions}
\label{sec: conclusion and future work}

This study conducts a driving simulator experiment to systematically examine how Situational Awareness (SA) and Spare Capacity (SC), in addition to baseline Driver Characteristics (DC), help capture takeover performance. We employ XGBoost models and SHAP values to evaluate the influence of significant factors on ten performance metrics. To ensure robustness, results are cross-validated using Random Forest and LightGBM models, and further compared with existing literature to support external validity. The results highlight the distinct yet complementary roles of SA and SC in shaping various performance dimensions: SA predominantly facilitates faster takeover responses (especially reflexive actions), while SC more substantially influences takeover quality - with lower SC levels generally correlating with poorer subjective comfort ratings and more abrupt control operations. 

The findings advance the scientific understanding of cognitive underpinnings in driver-vehicle interactions, providing empirical evidence that SA and SC influence separate but interrelated aspects of takeover behavior. By distinguishing these roles, the study contributes to a more refined theoretical framework for modeling driver performance during control transitions in conditionally automated driving. This differentiation is also practically significant: it informs the design of human-centered automated systems that can monitor and respond to drivers’ cognitive states in a targeted manner. Rather than relying on a single readiness indicator, future systems could optimize takeover support by tailoring alerts or interventions based on whether rapid response or quality of control is the priority. Taken together, these insights lay foundational knowledge for the development of more adaptive, personalized automation strategies. Future research should build on this foundation by examining the dynamic interplay between SA and SC, and by validating these findings in real-world or longitudinal studies to enhance validity and generalizability.

\section*{Declaration of Generative AI and AI-assisted technologies in the writing process}

During the preparation of this work the authors used ChatGPT in order to improve language and readability. After using this tool, the authors reviewed and edited the content as needed and take full responsibility for the content of the publication.

\bibliographystyle{cas-model2-names}
\bibliography{cas-refs}

\appendix

\renewcommand{\appendixname}{Appendix~Alph{section}}

\section*{APPENDIX}

\section{Driver characteristic questionnaire}
\label{appendix a: driver characteristic questionnaire}

This study employs the following questionnaire to assess driver characteristics, drawing from established instruments. The questionnaire collects drivers' background information (Table \ref{tab: background}), risk-taking attitudes (Table \ref{tab: risk-taking attitude}), trust in the conditionally automated driving systems (Table \ref{tab: trust}), takeover skills (Table \ref{tab: takeover skills}), and takeover styles (Table \ref{tab: takeover style}).



\begin{table*}[!h]
\centering
\caption{Background questions, based on \cite{nordhoff2023driver} and \cite{lu2017much}.}
\label{tab: background}
\resizebox{\textwidth}{!}{ 
\begin{tabular}{l l }

\toprule
\textbf{Latent Variables} &  \textbf{Observed Variables} \\ \midrule

$age$&What is your age?  \\
 
$gender$&What is your gender?  [Male; Female; Others]\\ 
$accu\!\_years$&How many years of driving experience do you have? \\
 
$accu\!\_dis$&How many kilometres (approximately) have you driven in the past 12 months? \\ 
$driving\!\_f\!re$&How many days (on average) have you driven per week in the past 12 months? \\ 

$driving\!\_skill$&How is your general driving skill? Please select the option that best matches your situation.\\
 & [Inexperienced; Intermediate; Experienced] \\

$assist\!\_f\!re$&How often (out of 10 times) do you use driver assistance functions while driving in the past 12 months?\\ 

\bottomrule

\end{tabular}}
\end{table*}


\begin{table*}[!h]
\centering
\caption{Scale measuring drivers' risk-taking attitude ($RT\!A$), based on \cite{ma2010safety}, \cite{taubman2004multidimensional}, and \cite{lajunen1995driving}.}
\label{tab: risk-taking attitude}
\begin{tabular}{c c p{36em} }

\toprule
\textbf{Latent Variables} & \textbf{Abbr.}& \textbf{Observed Variables}  \\ \midrule

\multirow{11}{*}{$RT\!A$}  & $RT\!A_{1}$&I follow the traffic rules most of the time $\left[ - \right]$.\\
                      & $RT\!A_{2}$&I drive cautiously most of the time $\left[ - \right]$ . \\
                      & $RT\!A_{3}$&I am not willing to compete with other drivers in traffic $\left[ - \right]$. \\
                      & $RT\!A_{4}$&I try to keep sufficient distances to the cars in front most of the time $\left[ - \right]$. \\
                      & $RT\!A_{5}$&I am willing to give up my right of way to other drivers to ensure safety $\left[ - \right]$. \\
                      & $RT\!A_{6}$&I enjoy the feeling of pushing a car to its maximum capability limits. \\
                      & $RT\!A_{7}$&It makes sense to exceed speed limits to get ahead of drivers who drive erratically, slowly, or extremely cautiously.\\
                      
                      & $RT\!A_{8}$&Engaging in risky driving behaviours does not necessarily mean someone is a bad driver. \\
                      & $RT\!A_{9}$&It's acceptable to break some traffic rules if they are restrictive. \\
                      & $RT\!A_{10}$&It's acceptable to drive at the moment when traffic lights change from yellow to red. \\

\bottomrule
\end{tabular}
\begin{tablenotes}
    \footnotesize
    \item $\left[ - \right]$ indicates reversed questions.
\end{tablenotes}

\end{table*}



\begin{table*}[!h]
\centering
\caption{Scale measuring drivers' trust in conditionally automated driving, based on \cite{nordhoff2021perceived}.}
\label{tab: trust}
    
\begin{tabular}{c c p{36em} }

\toprule
\textbf{Latent Variables} & \textbf{Abbr.} & \textbf{Observed Variables} \\ \midrule
\multirow{3}{*}{$trust$} 
                         & $T_{1}$ & I trust the automated car to maintain sufficient distances from the cars around me.\\
                         & $T_{2}$  & I trust the automated car to effectively detect the collisions ahead that it can not handle.\\
                         & $T_{3}$  & I trust the automated car to alert me to take over car control in time.\\ 

\bottomrule
\end{tabular}
\end{table*}



\begin{table*}[!h]
\centering
\caption{Takeover skill inventory, modified from Driver Skill Inventory \citep{lajunen1995driving}.}
\label{tab: takeover skills}
    
\resizebox{\textwidth}{!}{ 
\begin{tabular}{c c l }

\toprule
\textbf{Latent Variables} & \textbf{Abbr.}& \textbf{Observed Variables} \\ \midrule

\multirow{9}{*}{Perceptual-Motor Skills ($P\!M\!S$)}& $P\!M\!S_{1}$ & Taking over car control from automation fluently was easy for me.  \\
                                           & $P\!M\!S_{2}$ & Adjusting driving speed was easy for me.\\
                                           & $P\!M\!S_{3}$ & Controlling the car was easy for me.  \\
                                           & $P\!M\!S_{4}$ & Bypassing the detected collisions ahead was easy for me. \\
                                           & $P\!M\!S_{5}$ & I realized that I needed to take over car control from automation before the takeover requests.\\

                                           & $P\!M\!S_{6}$ & I reacted to takeover requests fast.   \\
                                           & $P\!M\!S_{7}$& I knew the right actions to take in response to the takeover requests. \\
                                           & $P\!M\!S_{8}$ & I made firm decisions to take over car control from automation. \\ \midrule

\multirow{8}{*}{Safety Skills ($S\!S$)} & $S\!S_{1}$ & I followed the traffic rules while taking over car control from automation.  \\
                                 & $S\!S_{2}$ & I was cautious while taking over car control from automation.  \\
                                 & $S\!S_{3}$ & I paid attention to the cars around me in automated mode.  \\
                                 & $S\!S_{4}$ & I paid attention to the cars around me while taking over car control.  \\
                                 & $S\!S_{5}$ & Keeping sufficient distances from the cars ahead was easy for me.  \\
                                 & $S\!S_{6}$ & Merging into the adjacent lane was easy for me.   \\
                                 & $S\!S_{7}$ & Braking effectively (i.e., not too hard nor too soft) was easy for me.  \\
                                 & $S\!S_{8}$ & I did not cause risks to myself and the cars around me.  \\ \bottomrule

\end{tabular}}

\end{table*}

\begin{table*}[!h]
\centering
\caption{Takeover style inventory, modified from the Multidimensional Driving Style Inventory \citep{taubman2004multidimensional}.}
\label{tab: takeover style}
\resizebox{\textwidth}{!}{ 
\begin{tabular}{c c l}

\toprule
\textbf{Latent Variables} & \textbf{Abbr.} & \textbf{Observed Variables} \\ \midrule

\multirow{5}{*}{$takeover\!\_st\!yle\!\_reckless$}& $RC_{1}$ & I misjudged the speed of the cars passing me when I was taking over car control.  \\
                                           & $RC_{2}$ & I forgot to switch on the turn indicator before changing lanes. \\
                                           & $RC_{3}$ & I nearly crashed due to misjudging my distances from other cars.  \\
                                           & $RC_{4}$ & I engaged in mind wandering from time to time when I was driving the car manually. \\
                                           & $RC_{5}$ & I tried to move into the left lane as soon as possible.  \\
                                           \midrule

\multirow{5}{*}{$takeover\!\_st\!yle\!\_anxious$} & $A_{1}$ & I felt nervous when I was taking over car control. \\
                                 & $A_{2}$ & Taking over car control frustrated me. \\
                                 & $A_{3}$ & It worried me when taking over car control after engaged in the n-back task.  \\
                                 & $A_{4}$ & I drove at or below the speed limit when I was taking over car control. \\
                                 & $A_{5}$ & I used muscle relaxation techniques (such as taking deep breaths).\\
                                 \midrule

\multirow{5}{*}{$takeover\!\_st\!yle\!\_angry$} & $A\!H_{1}$ & I swore at the automation when it asked me to take over car control. \\
                                 & $A\!H_{2}$ & I wanted to blow my horn or ``flash'' the car in front as a way of expressing frustrations.  \\
                                 & $A\!H_{3}$ & I enjoyed the excitement of taking risks when I was taking over car control.  \\
                                 & $A\!H_{4}$ & I took chances to merge into the adjacent lane. \\
                                 & $A\!H_{5}$ & I removed at least one hand from the steering wheel when I was driving the car manually.  \\ \midrule

\multirow{5}{*}{$takeover\!\_st\!yle\!\_patient$} & $PC_{1}$ & I waited for a proper gap to change lanes.\\
                                 & $PC_{2}$ & I based my takeover behaviours on the motto ``better safe than sorry''.  \\
                                 & $PC_{3}$ & I took over car control from automation cautiously. \\
                                 & $PC_{4}$ & I shifted my focus from the game to taking over car control before the takeover requests. \\
                                 & $PC_{5}$ & I had to slam on the brake to avoid collisions $\left[ - \right]$. \\
                                 \bottomrule

\end{tabular}}
\begin{tablenotes}
    \footnotesize
    \item $\left[ - \right]$ indicates reversed items.
\end{tablenotes}
\end{table*}

\section{Driver characteristic distribution}
\label{appendix b: driver distribution}

This appendix presents the descriptive statistics of participant demographics, skill-related attributes, and style-related traits in Table \ref{tab: driver characteristics}. All data were collected using the questionnaire detailed in Appendix \ref{appendix a: driver characteristic questionnaire}.

\begin{table*}[!h]
\centering
\caption{Descriptive statistics of participants' driver characteristics.}
\label{tab: driver characteristics}
\begin{tabular}{p{6em} p{3em}p{3em} p{2em} p{6em}  p{6em} p{3em}p{3em} p{2em} p{2em}}

\toprule
\textbf{Characteristics} &  \textbf{mean} &\textbf{SD} &  \textbf{min} &\textbf{max} & \textbf{Characteristics} & \textbf{mean} & \textbf{SD}&  \textbf{min} &\textbf{max} \\ \midrule
$age$ & 37.47 & 16.58& 18 & 72& $takeover\!\_skill$  &3.73  & 0.55 & 2.56& 4.94\\
$gender$ & 0.42 & 0.50 & 0 & 1& $RT\!A$  &2.18  & 0.49& 1& 3.2\\
$accu\!\_years$ & 17.04 & 16.91 & 1 & 55& $trust$  & 3.98& 0.88& 2& 5\\
$accu\!\_dis$  & 5840.91 & 6936.30& 0& 30000& $st\!yle\!\_reckless$  &2.57 & 0.73& 1&4.4 \\
$driving\!\_f\!re$  &0.30 & 0.27 & 0& 1& $st\!yle\!\_anxious$ & 2.55& 0.69& 1& 4.2	\\
$driving\!\_skill$  & 1.40 & 0.66& 0& 2& $st\!yle\!\_angry$  &  1.96& 0.60& 1& 3.2\\
$assist\!\_f\!re$ & 0.30&0.37& 0& 1& $st\!yle\!\_patient$  & 3.42& 0.49& 2.4 &4.4 \\
                                 \bottomrule
\end{tabular}									

\begin{tablenotes}
    \footnotesize
    \item $*$ For $gender$ : Male = 0; Female = 1.
    \item $*$ For $driving\!\_skill$: Inexperienced = 0; Intermediate = 1; Experienced = 2.
    \item $*$ Characteristics on the right side of the table (including $takeover\!\_skill$, $RT\!A$, $trust$, $st\!yle\!\_reckless$, $st\!yle\!\_anxious$, $st\!yle\!\_angry$, and $st\!yle\!\_patient$) are assessed on five-point scales ranging from 1 to 5.
\end{tablenotes}
\end{table*}

\section{Metric definition}
\label{appendix c: metric collection}

This Appendix details the definitions of Situational Awareness (SA)-related visual metrics (Table \ref{tab: sa visual metric definition}), Spare Capacity (SC)-related visual metrics (Table \ref{tab: sc visual metric definition}), and drivers' operational metrics (Table \ref{tab: operational metric}).

\begin{table*}[!h]
\centering
\caption{Definitions of Situational Awareness (SA)-related visual metrics during the 30 seconds preceding takeover requests.}
\label{tab: sa visual metric definition}
\renewcommand{\arraystretch}{1.5}
\begin{tabular}{p{5em}  p{3em} p{28em}  p{8em} }

\toprule
Metrics & Unit & Description & Reference\\ \midrule
$nr_{road}$ & -- & the number of fixations on the forward road & \multirow{13}{8em}{\cite{kastle2024objective, zhou2021using}}\\ 
$nr_{rearview}$ &-- & the number of fixations on the rear-view mirror  &\\
$nr_{sideview}$&--& the number of fixations on the left-wing / right-wing mirrors  &\\ 

$nr_{H\!M\!I}$& -- & the number of fixations on the text alarm / the takeover button  &\\
$nr_{dashboard}$& --& the number of fixations on the dashboard  &\\
$f_{max}$& $s$ & maximum fixation time  &\\
$f_{mean}$ & $s$ &  mean fixation time  &\\
$f_{std}$& $s$ &  standard deviation of fixation time  &\\
$num_{f}$ & --&  overall number of fixations  &\\
$s_{max}$ &$px/s$&  maximum saccade velocity   &\\
$s_{mean}$ &$px/s$&  mean saccade velocity  &\\
$s_{std}$ &$px/s$&  standard deviation of saccade velocity  &\\
$num_{s}$ & -- &  the overall number of saccades  &\\

\bottomrule
\end{tabular}
\end{table*}

\begin{table*}[!h]
\centering
\caption{Definitions of Spare Capacity (SC)-related visual metrics for the period between the initiation of takeover requests and the completion of lane changes.}
\label{tab: sc visual metric definition}
\renewcommand{\arraystretch}{1.5}
\begin{tabular}{ p{3em} l p{8em} p{26em} p{6em} }

\toprule
 Metrics &Unit & Description & & Reference\\ \midrule
$t_{road}$ & $s$&\multirow{4}{9em}{the interval between the initiation of a takeover request and the moment ...} &  the driver first establishes visual fixation on the forward road & \multirow{16}{6em}{\cite{deniel2023gaze,liang2021using,sharma2024review}}\\ 
 $t_{H\!M\!I}$  &$s$& &the driver first establishes visual fixation on the text alarm / the takeover button & \\
 $t_{mirror}$ &$s$& &the driver first establishes visual fixation on the left-wing / right-wing mirrors & \\ \cline{1-4}

 $DU\!R_{road}$ & s& \multirow{3}{9em}{the total time duration that ...}& the driver fixates on the forward road& \\
 $DU\!R_{H\!M\!I}$ &s & & the driver fixates on the text alarm / the takeover button & \\
 $DU\!R_{mirror}$ & s & & the driver fixates on the left-wing / right-wing mirrors & \\ \cline{1-4}

 $N\!O_{road}$ & -- &\multirow{3}{9em}{the number of fixations on ...}& the forward road& \\
 $N\!O_{H\!M\!I}$ & -- & & the text alarm / the takeover button & \\
 $N\!O_{mirror}$ & -- && the left-wing / right-wing mirrors & \\ \cline{1-4}

 $AV\!G_{road}$ &$s$& \multirow{3}{9em}{the average duration of each fixation on ...}& the forward road& \\
 $AV\!G_{H\!M\!I}$ &$s$&& the text alarm / the takeover button & \\
 $AV\!G_{mirror}$ &$s$&& the left-wing / right-wing mirrors & \\ 

\bottomrule
\end{tabular}
\end{table*}

\begin{table*}[!h]
\centering
\caption{Definitions of drivers' operational metrics during takeovers.}
\label{tab: operational metric}
\renewcommand{\arraystretch}{1.5}
\resizebox{\textwidth}{!}{    
\begin{tabular}{  p{3em} l p{8em} p{26em} p{8em} }

\toprule
 Metrics & Unit & Description & & Reference\\ \midrule
 $t_{buttion}$ & s &  \multirow{3}{9em}{the interval between the initiation of a takeover request and the moment ...} & the driver presses a button to activate manual vehicle control & \multirow{4}{9em}{\cite{liang2024examining, miller2015distraction, gold2013take}}\\ 
$t_{steering}$ & s & &the steering wheel angle exceeds 2 degrees & \\
$t_{pedal}$ & s & &the braking/accelerator pedal position surpasses 10\% & \\ \cline{1-4}

$T\!oT$  &s & \multicolumn{2}{l}{the shorter time between $t_{steering}$ and $t_{pedal}$} &\\
\midrule
$TTC$  &s & \multicolumn{2}{l}{the remaining time before a collision occurs if the current course and speed difference remain unchanged} & \cite{hyden1996traffic}\\

\bottomrule
\end{tabular}
}
\end{table*}

\end{document}